\documentstyle[12pt,aaspp4,tighten]{article}
\def\etal{{\it et al.~}}

\def\arcsec{{\prime\prime}}

\def\VEV#1{\left\langle #1\right\rangle}
\def\abso#1{\mid\! #1\!\mid}
\def\la{\hbox{ \raise.35ex\rlap{$<$}\lower.6ex\hbox{$\sim$}\ }}
\def\ga{\hbox{ \raise.35ex\rlap{$>$}\lower.6ex\hbox{$\sim$}\ }}

\slugcomment{\vbox{\hbox{CAL-637}}}
\begin{document}
\title{Constraining $\Omega_0$ With The Angular-Size Redshift Relation Of
Double-Lobed Quasars In The FIRST Survey}

\author{Ari Buchalter$^*$\altaffilmark{1}, David J. Helfand$^*$\altaffilmark{2},
Robert H. Becker$^\dagger$\altaffilmark{3}, Richard L. White
$^\ddagger$\altaffilmark{4}}
\affil{$^*$Department of Astronomy, Columbia University, New
York, NY 10027}
\affil{$^\dagger$Department of Physics, University of California, Davis, CA
 95616 and IGPP/Lawrence Livermore National Laboratory}
\affil{$^\ddagger$Space Telescope Science Institute, 3700 San Martin Dr.,
 Baltimore, MD 21218}
\altaffiltext{1}{ari@astro.columbia.edu}
\altaffiltext{2}{djh@astro.columbia.edu}
\altaffiltext{3}{bob@igpp.llnl.gov}
\altaffiltext{4}{rlw@stsci.edu}
%\altaffiltext{5}{ctw@astro.columbia.edu}

\begin{abstract}

In previous attempts to measure cosmological parameters from the angular size-redshift ($\theta$-$z$)
relation of double-lobed radio sources, the observed data have generally been consistent with a 
static Euclidean
universe, rather than with standard Friedmann models, and past authors have 
disagreed significantly as to what effects are responsible for this observation.
These results and different interpretations may be due largely to a variety of selection effects 
and differences in the sample definitions destroying the integrity of the data sets, 
and inconsistencies in the analysis undermining the results. 
Using the VLA FIRST
survey, we investigate the $\theta$-$z$ relation for 
a new sample of double-lobed 
quasars. We define a set of 103 sources, carefully 
addressing the various potential problems which, we believe, have compromised
past work, including a robust definition of size and 
the completeness and homogeneity of the sample, 
and further devise a self-consistent method to 
assure accurate morphological classification
and account for finite resolution effects in the analysis.
Before focusing on cosmological constraints, we investigate the possible impact of correlations
among the intrinsic properties of these sources over the entire assumed range of allowed cosmological 
parameter values. For all cases, we find apparent size evolution of the form $l \propto (1+z)^c$
with $c \approx -0.8 \pm 0.4$ which is found to arise mainly from a power-size correlation
of the form $l \propto P^{\beta}$ ($\beta \approx -0.13
\pm 0.06$) coupled with a power-redshift correlation. Intrinsic size evolution is consistent
with zero. We also find that in all cases, a subsample with $c \approx 0$ can be defined, whose 
$\theta$-$z$ relation should therefore arise primarily from cosmological effects. These results
are found to be independent of orientation effects, though other evidence indicates that
orientation effects are present and consistent with predictions of the unified scheme for radio-loud active
galactic nuclei. The above results are all confirmed by non-parametric analysis.

Contrary to past work, we find that the observed $\theta$-$z$ relation for our sample
is more consistent with standard
Friedmann models than with a static Euclidean universe. Though the current data
cannot distinguish with high significance between various Friedmann models, 
significant constraints on the cosmological parameters within a given model are obtained. 
In particular,
we find that a flat, matter-dominated universe
($\Omega_0 = 1$),
a flat universe with a cosmological constant, and an open universe all provide comparably good fits
to the data, with the latter two models both yielding $\Omega_0 \approx 0.35$ with 1$\sigma$ ranges
including values between $\sim 0.25$ and 1.0; the $c \approx 0$ subsamples yield values of $\Omega_0$
near unity in these models, though with even greater error ranges. We also examine the values
of $H_0$ implied by the data, using plausible assumptions about the intrinsic source sizes, and find 
these to be consistent
with the currently accepted range of values. We determine the sample size needed to
improve significantly the results, and outline future strategies for such work.

\end{abstract}


\section{INTRODUCTION}

The angular size-redshift ($\theta$-$z$) relation for a cosmological population
of standard rods is a powerful probe of the large-scale geometry of the 
universe. For a universe characterized by the Friedmann-Robertson-Walker
metric, with curvature arising from the energy density of ordinary matter, and 
possibly a cosmological constant, the angular size of a rod of intrinsic length
$l$, viewed at an angle $\phi$ to the line of sight is given by
\begin{equation}
\theta={{l \sin\phi} \over D_A}; \qquad D_A= {cR_{0} \over {1+z}}
\sum_{}^{} \left( \int_{1}^{1+z} \frac {dx} {H_{0}R_{0}\left(
\Omega_{0}x^3 + (1-\Omega_{0}-\Omega_{\Lambda})x^2 + \Omega_{\Lambda}
 \right)^{1/2}} \right) 
\label{angular size}
\end{equation}
(Weinberg 1972) where $D_A$ is the angular-size distance, 
$H_0$ is the present value of the Hubble constant, $R_0$ is the expansion
scale factor in units of time, $c$ is the speed of light,
$\Omega_{0}$ is the present ratio of the matter density to the critical 
density, $\Omega_{\Lambda} = \Lambda/3{H_{0}^2}$
(where $\Lambda$ is the cosmological constant), and $\Sigma(x) = \sin{x},{x},\sinh{x}$
for closed, flat, and open geometries, respectively. Contributions to the energy
density arising from more exotic phenomena such as textures will affect the
angular size in a straightforward manner, but are not considered here.
Figure 1 illustrates the $\theta$-$z$ relation for deprojected rods ($\phi=90^{\circ}$)
with an intrinsic size
of $l = 200h_{0}^{-1}$ kpc ($h_0$ is the Hubble constant in units of 100 km s$^{-1}$ 
Mpc$^{-1}$) for three Friedmann cosmologies: 1) an Einstein de-Sitter 
universe, with $\Omega_{0} \equiv 1$ and $\Omega_{\Lambda}=0$, 2) a flat
universe with $0 < \Omega_{0} \leq 1$, $\Omega_{0} + \Omega_{\Lambda} = 1$, 
and 3) a non-closed, matter-dominated universe with $0 < \Omega_{0} \leq 1$, 
$\Omega_{\Lambda} = 0$
(we will hereafter refer to these as models 1, 2, and 3, 
respectively). Model 1 is, in effect, a limiting case of models 2 and 3,
and we do not consider models with $\Omega_0 > 1$. 
The particular values chosen for $\Omega_{0}$ and 
$\Omega_{\Lambda}$ in the Figure are listed, and
the curve for a static Euclidean universe, with $\theta \propto 
z^{-1}$ (hereafter referred to simply as the Euclidean model), 
is shown for comparison. The amplitudes of the Friedmann curves are scaled 
by $h_0$, while their
shapes and, in particular, the location of the minimum in the angular size
(typically between $z=1$ and $z=2$),
depend on $\Omega_{0}$ and $\Omega_{\Lambda}$. For randomly
oriented rods, these curves actually define upper limits to the observed
angular size distribution, since projection effects will scatter the 
observed sizes downward. Note that for the conventional parameter values 
listed in Figure 1,
the curves for the different Friedmann models, particularly the underdense
($\Omega_0$ possibly $< 1$) models 2 and 3, are fairly similar.
\placefigure{fig1}

With the discovery of double-lobed radio galaxies and quasars,
it was hoped that a standard rod had been found which could 
constitute a useful high-redshift
sample for measuring the geometry of the universe. Early work 
(Miley 1971; Wardle \& Miley 1974; Hooley \etal 1978)
revealed, however, that
the upper limit to the $\theta$-$z$ data traced a Euclidean curve,
implying that some effect must be diminishing the apparent sizes of these
objects at high redshift. Subsequent studies (Kapahi 1985; Singal 1988; Barthel
\& Miley 1988; Kapahi 1989; Onuora 1989; 
Ubachukwu \& Onuora 1993; Nilsson \etal 1993; Chy\.{z}y \& Zi\c{e}ba 1993, Singal 1993), 
looking at the 
variation of binned mean or median angular sizes as a function of 
redshift to compensate
for projection effects, all confirmed that the observed $\theta$-$z$ data
was strikingly consistent with a Euclidean model. 

Three main explanations have been proposed
to account for this observation: 1) The characteristic length
scale of double-lobed sources may change with cosmic epoch,
presumably due to differences in the density of the intergalactic medium (IGM)
and/or changes in the energetics of the active galactic nuclei (AGN) which
power these sources. Intrinsic size evolution of the form 
$l \propto (1+z)^{n}$, with $n<0$ so that higher redshift objects are 
intrinsically smaller, would reconcile the data with Friedmann models.
2) Since power, $P$, and redshift are necessarily 
correlated in any flux-limited sample, usually approximated by $P \propto 
(1+z)^{x}$, a negative 
correlation between power and intrinsic size, typically parameterized
as $l \propto P^{\beta}$, with $\beta < 0$, would give rise to an apparent correlation
between $l$ and $z$, given by $l \propto (1+z)^{{\beta}{x}}$, 
effectively mimicing size evolution. In general, effects 1) and 2) may both be
present, giving rise to overall observed size evolution of the form $(1+z)^c$, where
$c=\beta x + n$ (Ubachukwu 1995).
3) According to the 
unified model for radio sources, the classification of an object as a radio 
galaxy or a quasar depends
only on its orientation, with quasars having inclinations within about 
$45^{\circ}$
of the line of sight, with a median inclination of $31^{\circ}$,
and radio galaxies being inclined roughly 
between $45^{\circ}$ and $90^{\circ}$, with a median inclination
of $69^{\circ}$ (Barthel 1989; Lister \etal 1994). 
If the unified scheme is correct, then in studies which include both radio 
galaxies and quasars, the high-redshift population (dominated by quasars)
would have systematically smaller mean angular sizes than the low-redshift
population (dominated by radio galaxies), making the universe appear
more Euclidean. A similar scheme has been proposed
to unify the two classes of radio-loud quasars, the core-dominated and
lobe-dominated quasars (CDQs and LDQs) (Orr \& Browne 1982), assuming
that only the radio flux from the compact core is relativistically beamed. In this model,
the moderately beamed LDQs, with a median ratio of the core-to-lobe flux density, $R$ 
$\sim 0.1$, have a median inclination of $40^{\circ}$ to the line of sight, while
the more strongly beamed CDQs, with a median $R$ 
$\sim 10$, have a median inclination of $10^{\circ}$ (Hough \& Readhead 1989; 
Ubachukwu 1996). If the observed fraction of CDQs increases with redshift,
as might be expected in a flux-limited sample, then even studies limited to quasar 
samples would also reveal a deficit of larger sources at higher redshifts. 

While most previous studies agree that the observed data follow a Euclidean
trend, they disagree substantially as to what combination of the above effects
is responsible. Several authors find evidence for significant size evolution
(Kapahi 1985; Oort \etal 1987; Barthel \& Miley 1988;  
Kapahi 1989; Neeser \etal 1995), while
others claim to find an $l$-$P$ correlation with little or no 
intrinsic size evolution
(Hooley \etal 1978; Masson 1980; Onuora 1991; Chy\.{z}y \& Zi\c{e}ba 1993; 
Nilsson \etal 1993). Moreover, there has been considerable disagreement as to 
whether the properties of radio galaxies and quasars follow similar trends. 
Some authors find evidence for stronger size evolution in the double-lobed radio
galaxy population (Onuora 1989; Chy\.{z}y \& Zi\c{e}ba 1993; Singal 1993) than in the
double-lobed quasar population, while several workers
have found a negative $l$-$P$ correlation for quasars but a positive one for
radio galaxies (Chy\.{z}y \& Zi\c{e}ba 1993; Nilsson \etal 1993; Singal 1993). 
Others have claimed
to see identical trends in the two populations (Gopal-Krishna \& Kulkarni 1992)
and even to reconcile the observed $\theta$-$z$
data with Friedmann cosmologies based on 
orientation effects within the unified scheme (Onuora 1991).

The lack of concordance among these previous results strongly suggests that the
construction and analysis of samples of double-lobed objects has been dominated by
systematic and/or selection effects which are
unrelated to the intrinsic behavior of the sources. In fact, several
investigations have traced the inconsistent results of various studies to such 
selection effects (Neeser \etal 1995) or to different sample definitions 
(Nilsson \etal 1993). We believe that other substantive issues
have not been properly incorporated
into the study of the $\theta$-$z$ relation for double-lobed sources and have
also compromised the results. We summarize these issues below:

$\;$ 1. In determining the morphological properties of double-lobed sources, 
it is desirable to characterize each source using
parameters that make no {\em a priori} assumptions about the source structure and are related
as directly as possible to the measured data. The moments of the brightness 
distribution, $B(\mbox{\boldmath $\vartheta$})$, where $\mbox{\boldmath $\vartheta$}$ 
represents two-dimensional, quasi-Cartesian coordinates, form one such set of parameters (Burn \& Conway 
1976). In particular, the second moment, 
$\theta_{sm}=2 \left[\int{ \mbox{\boldmath $\vartheta$}^{2} B(\mbox{\boldmath $\vartheta$}) 
d\mbox{\boldmath $\vartheta$}}/ \int
{B(\mbox{\boldmath $\vartheta$}) d \mbox{\boldmath $\vartheta$}} \right]^{1/2}$, is a model-independent measure of the source
size which, for sufficiently large sources (\ga 1/3 of the beam), 
is independent of the beam resolution (Condon 1988; Coleman 1996).
For double-lobed sources, however, $\theta_{sm}$ is an unstable diagnostic since it is a 
flux-weighted quantity; two sources with identical lobe structures (i.e., apparent
shapes, sizes, and lobe-lobe separation) would 
yield a different value of $\theta_{sm}$ a) if one exhibited a significant core component
(e.g., because it was an intrinsically large source with its core flux relativistically boosted
by projection, as opposed to a smaller, de-projected source with the same apparent lobe properties
but no detected core flux),
b) if the ratio of the lobe fluxes in the two sources differed, or c) if the maps of the two 
source fields had different signal-to-noise properties. 
In addition, surveys with different flux 
sensitivities and limiting resolutions can yield different values of $\theta_{sm}$
for the same source.
The more commonly used measure of the angular size of double-lobed
sources is the
``largest angular size'' (LAS), typically taken to be either the maximum linear extent
over which a given level of radio emission is detected, $\theta_{max}$, 
or the peak-to-peak angular
separation, $\theta_{pp}$. The former definition, however, is also a poor measure of size
since it is highly sensitive to the details of the observation; 
radio observations 
conducted with different instruments at different 
frequencies, with different flux sensitivities and beam widths, can yield 
drastically different values of $\theta_{max}$ for the same object. An example of this 
is illustrated in Neeser \etal (1993). Even within a given radio data set, the
measured value of $\theta_{max}$ for an arbitrary distribution of
high-redshift, extended objects is highly susceptible to the effects of
cosmological surface brightness dimming. 

For these reasons, many authors studying the 
$\theta$-$z$
relation for double-lobed sources focus on Fanaroff-Riley type II (FR-II) objects
(Fanaroff \& Riley 1974), which exhibit radio-bright hot-spots near the outer edges
of the lobes. For these objects, the peak-to-peak size is largely independent
of the details of the observation, and thus provides
a fairly robust measure of the angular size (see \S2 for further discussion)
\footnote[1]{Though $\theta_{pp}$ is fairly insensitive to details of the
observation, it may be asked whether the hot-spots in different FR-II sources occur
at the same relative positions, so that one is in fact measuring a stable quantity for different sources.
The canonical criterion for FR-II sources is that the ratio, $FR=\theta_{pp}/\theta_{max}$, 
where $\theta_{max}$ is defined as the greatest linear extent of the outer lobes measured to the 1\% contour, 
be greater than
0.5. Though $\theta_{max}$ may generally depend on the details of the observation,
studies have shown that FR-IIs invariably tend to have FR values near unity.
Rector, Stocke \& Ellingson (1995) study a sample of 30 FR-IIs in the range
$0.26 < z < 0.63$ and find, for those
with well determined values of FR, a mean FR value of 0.85 with a standard deviation of 0.07,
with all having FR $> 0.7$. Thus, the relative peak-to-peak scales in these sources vary
by at most 17\% from the mean value, though typically much less, and show no systematic variation
with redshift. It should be noted that
the spread in FR may arise from the fact that if the lobes themselves
are not spherically symmetric,
FR will vary simply because the apparent location of the hot-spots within the optically thin
lobes will vary for sources with different projection angles, even if the relative positions of the
hot spots within the lobes of these sources are identical, suggesting that $\theta_{pp}$ is in fact
a more stable quantity that FR. The variation in FR may also be due 
in part to instrumental effects and/or cosmological surface brightness dimming
operating in the determination of
$\theta_{max}$, rather than intrinsic differences in $\theta_{pp}$.}.
However, the peak positions, and thus peak-to-peak sizes, are
typically derived from fitting
a specific model (usually Gaussians) to the brightness distribution
of each source. Fitting Gaussians to the highly asymmetric, edge-brightened lobe components 
of FR-IIs
will necessarily return peak positions which are slightly closer to the central core,
and thus {\em underestimate} the peak-to-peak angular size. The resulting fractional
error in the angular size would be small for large sources, but may be appreciable
for smaller sources where the resolved lobe size is comparable to the angular
distance between lobes. For small enough sources, standard Gaussian fitting routines
may fit only one or two components to a source which clearly exhibits a more complicated
morphology, and thus drastically underestimate $\theta_{pp}$.
These effects would become most pronounced at
higher redshifts ($0.5\,\la\,z\,\la\,3.0$),
where the relative fraction of smaller sources is greatest (cf. Figure 1),
and precisely where cosmological effects in the $\theta$-$z$ plane become important,
thus making the universe appear more Euclidean at higher redshifts.
It is preferable therefore to measure $\theta_{pp}$ for FR-II sources directly from the
radio data of a single, high-resolution survey,
rather than from multiple survey catalogs generated by model-specific fitting algorithms. 
Despite these considerations, most of the $\theta$-$z$ studies to date have,
in fact, employed samples compiled from output catalogs of multiple radio surveys
(Miley 1971; Hooley \etal 1978; Singal 1988; 
Kapahi 1989; Nilsson \etal 1993; Singal 1993; Neeser \etal 1995).
A further danger in using data taken from multiple surveys is that such a study
may selectively
omit some sources altogether. For example, large, low-surface-brightness objects
detected in a low-resolution sample may be resolved out in a deep, 
high-resolution sample, and a high-frequency sample will 
generally contain more compact
sources than a low-frequency sample. The mixing of sources from different 
samples may thus destroy the consistency of the set required to measure the 
$\theta$-$z$ relation.

$\;$ 2. The sizes of double-lobed sources are measured from radio data, but 
the redshift information is often obtained in an optically selected fashion.
Thus, constructing a $\theta$-$z$ diagram from quasar catalogs (as all previous
studies have done) might mix radio-measured angular sizes with optically 
selected redshifts, and may introduce serious selection effects.
Though one may hope that the highly heterogeneous manner in which lists of 
quasar redshifts have been developed would ``wash out'' any such effects, it is 
easy to conceive of scenarios wherein the sizes of an optically selected
subset of double-lobed radio objects
could be systematically larger or smaller than the population as a whole. Such
a selection effect, if present, would operate even when using catalogs
with complete optical identifications and redshift information.
To eliminate this potential problem would require obtaining redshift information
for a complete set of {\em radio-selected} double-lobed objects, but such a 
$\theta$-$z$ study has not been carried out to date.

$\;$ 3. For the purposes of studying cosmology using the $\theta$-$z$ relation,
double-lobed sources are chosen to the extent that they might represent
a population of standard rods. It is clear, however, that a given 
double-lobed source does not maintain a fixed size, but
rathers grows with time, over a period of roughly $10^{7}$--$10^8$ yr 
(Gopal-Krishna \etal 1996) (This growth is not
to be confused with the intrinsic size evolution discussed above, which occurs
over cosmological time frames and refers to evolution of the overall length
scale characterizing these objects and not the growth in size of a given 
source). Thus, when considering the angular sizes of these sources, those
with smaller sizes will in general be a mixture of intrinsically smaller sources
plus larger sources viewed in projection. Moreover,
many radio sources
exhibit a core-jet structure which, if not well-resolved,
can easily be mistaken for a small double-lobed
morphology, and many small, double-lobed sources may not be sufficiently
resolved for an accurate classification. 
To avoid potential confusion, it is necessary to determine,
for a given set of radio observations,
the angular scale above which morphological classifications can be accurately 
determined. This scale will generally be significantly greater than the 
survey resolution limit. 
However, previous $\theta$-$z$ 
studies include angular sizes down to the
survey resolution limits (Barthel \& Miley 1988; Singal 1988;
Singal 1993; Nilsson \etal 1993; Neeser \etal 1995), and are thus 
mixing true double-lobed sources of various sizes 
together with objects which may in fact be a different class of sources,
and by using multiple surveys, are doing so in a highly non-uniform fashion.
The resulting admixture of objects is not likely to yield a good approximation to a standard
rod, and while interesting from the viewpoint of AGN evolution, is not valid 
for cosmological studies. 
Since the morphologies of objects with large angular sizes
are less likely to be misclassified, this problem also becomes more severe at higher
redshifts, where there is a greater fraction of smaller, less well-resolved sources.
The wrongful inclusion of more smaller sources at higher redshifts would
significantly {\em decrease} the mean angular size at these redshifts, again making
the universe appear more Euclidean.

$\;$ 4. Another important consequence of finite survey resolution
is that a constant angular resolution limit does not correspond to a constant
minimum linear size, but rather one which varies with redshift (cf. Eq.~(\ref{angular size})
and Figure 1). Thus, if all the $\theta$-$z$ data down to some 
limiting resolution are used,
the resulting sample will not only mix sources with different 
intrinsic sizes (due to the spread in the intrinsic size distribution),
but will span different ranges of intrinsic sizes
at different redshifts.  
This redshift-dependence of the intrinsic size distribution
again undermines the consistency needed to define a cosmological population of standard
rods.

The VLA FIRST Survey (Becker, White, \& Helfand 
1995) is the most sensitive survey of its kind, and
represents a valuable new tool for studying the $\theta$-$z$
relation. To date, the project has mapped $\sim 3,000$ square
degrees of the north Galactic cap at 1.4 GHz to a sensitivity of $\sim 1$ mJy with a 
$5.4^{\prime\prime}$
FWHM Gaussian beam, and has catalogued roughly 270,000 sources with subarcsecond
positional accuracy.
In this paper we investigate the $\theta$-$z$ relation for double-lobed quasars
in the FIRST survey. We construct a sample optimally suited
for studying the $\theta$-$z$ relation,
addressing the various problems and selection effects which may be present,
and devise analytic methods to account for these 
(\S2). In Section 3, we 
explore the relationships among the intrinsic properties of the sources,
using both parametric and non-parametric methods,
and find evidence, regardless of the chosen cosmological parameters,
for a negative correlation between power and size, with $\beta \approx -0.13 \pm 0.06$ 
which, coupled with
the observed power-redshift correlation, gives rise to apparent size evolution with
$c \approx -0.8 \pm 0.4$; 
intrinsic size evolution is consistent with zero. We also find that a subsample
can be defined for which $c$ is consistent with zero, implying that any observed
$\theta$-$z$ relation would be entirely due to cosmological effects. 
We find these results to be independent of orientation effects, though other
evidence confirms that orientation effects are present and consistent with the 
predictions of the unified scheme for radio-loud AGNs. 
In Section 4, we investigate the constraints that can be placed on cosmological parameters 
from the $\theta$-$z$ data and find that, contrary to past work,
the observed data are less consistent with a Euclidean
model than with standard Friedmann models. 
Both underdense models yield $\Omega_0 \approx 0.35$, with
1$\sigma$ intervals ranging from $\sim 0.25$ to 1.0, and the $c \approx 0$ 
subsample favors values near unity in these
models. Model 1, with $\Omega_0 \equiv 1$, provides a comparably good fit,
and even appears slightly favored by the $c \approx 0$ 
subsample, though these results are likely due to the reduced number of free parameters in this model.
Though all three Friedmann models yield consistent values of $\Omega_0$, the
data at present cannot distinguish between different 
cosmological models with reasonable
significance. 
As a consistency check on our analysis, we also investigate the values of $H_0$ implied
by the data, based on assumptions about the intrinsic sizes of the sources, and find
the results to be consistent with the range spanned by current estimates.
In Section 5 we present our conclusions and discuss future prospects for such work.
 
\section{THE SAMPLE}

\subsection{Selection Criteria}

In the interest of maintaining a maximally homogeneous population of objects,
we restrict our study of double-lobed sources to those identified as quasars,
excluding objects classified as radio galaxies. We thus bypass the issue
of whether the potentially different characteristics of these two kinds of objects, 
such as different mean orientations or power-size correlations, produce
non-cosmological effects in the $\theta$-$z$ plane. Moreover, since
all Friedmann models approach a Euclidean universe as $z \rightarrow 0$
(see Figure 1), the 
low-redshift population (dominated by radio galaxies) carries less
information about the cosmology. Conversely, since many quasars are found at
higher redshifts ($z \ga 1$), where the predictions of different models 
exhibit different minima and begin to diverge significantly, 
their $\theta$-$z$ distribution
is more sensitive to $\Omega_{0}$ and $\Omega_{\Lambda}$.
In addition, we further restricted our sample to sources with $z > 0.3$,
since, as mentioned in \S1, past work indicates that the properties of low-$z$
and high-$z$ radio sources exhibit different behaviors, with the cutoff occurring
roughly at $z=0.3$, beyond which quasars begin to dominate (Heckman \etal 1992; Hes \etal 1995).

Using the Hewitt \& Burbidge (1993), Veron-Cetty \& Veron (1996) and 
FIRST Bright QSO Survey (Gregg \etal 1996; Becker \etal 1997) catalogs, we selected all $z > 0.3$
quasars whose positions fell within the currently mapped area of the 
FIRST survey. Each of these
was then inspected separately by A.B. and C.T.W. \footnote[2]{We thank
Chelsea T. Wald for her work in analyzing the radio images.}  
on the FIRST radio maps to determine the radio morphology.
We included in our sample only
those sources where the quasar position fell near (i.e., within a few arcseconds of) 
the center of an edge-brightened, double-lobed radio source.
Restricting the study to edge-brightened, FR-II objects offers two main 
advantages. First, since the lobes have radio-bright hot spots,
the measured peak-to-peak angular sizes are less sensitive to instrumental
effects and cosmological surface brightness dimming (Neeser \etal 1995)
than are FR-I sources, 
whose lobe components fade gradually toward the edges. Second, since
the underlying mechanism that distinguishes FR-I and FR-II sources is not
well understood, restricting the analysis to a single type of object 
assures
us that the $\theta$-$z$ data are not altered by non-cosmological effects which
may arise from intrinsic differences between these objects. Determining the 
radio morphology (i.e., FR-I, FR-II, core-jet, etc.) 
is relatively easy for sources significantly larger than the beam, but can
be problematic for smaller sources,
either because of the limited survey resolution, or because uncertainties in
the optical positions of the quasars make it difficult to determine whether
the radio peak corresponds to a core or a lobe. It is unlikely that
FR-I sources would be misclassified as FR-IIs; cosmological surface brightness
dimming would select against very faint FR-Is
with $z > 0.3$, and recent studies have
shown that there is a real decline in the incidence of powerful
FR-Is with redshift (with
few having $z > 0.3$), while the number of FR-IIs increases with redshift
(Zirbel 1996). Following Neeser \etal (1995),
we classify smaller sources as FR-II only if the quasar position is at the 
center of two comparable edge-brightened lobes, excluding sources which 
suggested a core-jet morphology, such as those where the quasar position
falls much closer to one of the components, or those where one component is 
much brighter than the other.
Sources too small for an accurate
morphological classification were omitted, thus introducing an effective
size cutoff in the data, the significance of which is described in \S2.3.
We excluded from our sample any sources with highly distorted or bent morphologies,
whose apparent sizes have been severely influenced by
asymmetric interaction with the IGM. 

\subsection{Properties of the Sample}

A total of 103 objects satisfied our selection criteria. 
For each object we measured 
the peak-to-peak angular size, hereafter denoted simply by $\theta$, directly from the FIRST radio maps,
so as to avoid any systematic underestimation of the sizes which could arise from using
the peaks of the Gaussian model fits. We also recorded the 
FIRST flux density for each source component; for components which are well-resolved,
such as the compact core components or sufficiently small lobe
components, the measured FIRST fluxes are reliable.
However, due to its high resolution, the FIRST survey is prone to resolving out flux
from extended sources, such as the radio lobes of larger FR-IIs. 
Thus we also measured, where possible, the 1.4 GHz flux densities for our sources from
the NVSS radio survey (Condon \etal 1996), which is currently mapping 82\% of the
celestial sphere at 1.4 GHz to a sensitivity of 2.5 mJy with $45^{\arcsec}$ FWHM resolution.
Any difference between the FIRST and NVSS fluxes for a given source was ascribed to
the extended lobe components, so that the total flux density, $S_t$, corresponds to the value
measured by the NVSS survey (if available), while the core flux density, if any, 
is determined by the FIRST measurement. Many sources do not have a core component in the FIRST
catalog, either because none was detected, or, in the case of smaller sources,
because the Gaussian fitting algorithm did not generate a separate component for the core.
The data for our sample are listed in Table 1.
For comparative purposes, we also measured $\theta_{sm}$ (not shown) for each source using both the
FIRST and NVSS radio data.
\placetable{tab1}

Figure 2 shows a scatter plot of the $\theta$-$z$ data.
The errors in the measured values of $\theta$ are typically $\sim 1^{\arcsec}$,
far less than the scatter in the angular sizes at any redshift.
For graphical purposes, we bin the data in redshift, both in equal intervals
of $(1+z)^{-3/2}$ (which corresponds to equal time per bin in an Einstein-de Sitter universe) and
with roughly equal numbers per bin, and calculate both  
the mean values, $\VEV{\theta}$, and median values, $\theta_{med}$, together with the standard
errors of the mean values, and median absolute deviations
\footnote[3]{The median is the value about which the sum of the absolute deviations is minimized,
just as the mean defines the value about which the rms deviation is minimized.},
respectively, for each bin.
The results are shown in Figures 3a--d, along with the curves
from Figure 1, whose amplitudes (corresponding to the mean or median intrinsic
sizes) have been scaled to provide a rough visual fit.
Before turning to a formal discussion of the results, we describe 
several important features
of the data, and address in detail the aforementioned issues associated with
properly defining and analyzing such a sample.
\placefigure{fig2}
\placefigure{fig3}

$\;$ 1. The most striking feature of the data is that, 
regardless of the binning details,
the observed data seem to be more consistent with Friedmann models than 
with a Euclidean model. The 
Friedmann curves shown are not the best-fit results, but are merely intended 
for qualitative reference.
We defer the discussion of the best-fit 
values (\S4) until after we have addressed the properties and analysis of the
sample, including potential problems.
It can immediately be seen, however, that while the data are generally consistent
with curvature models, 
it is unlikely, given the uncertainties, that the current sample can distinguish
with high significance between the different models. Note that in each case, the data
point of the first bin appears anomalously high.
 
$\;$ 2. Since the radio data are derived entirely from one consistent data
set (with a single flux limit, beam width, and frequency),
our sample does not suffer from the potential problems, described in \S{1},
associated with the mixing of different samples.  

$\;$ 3. As expected, we find that for cases where the FIRST and NVSS flux measurements agree, 
the second moments derived from both survey maps 
generally agree to within the second moment errors (provided that the 
source was larger about 1/3 the NVSS beam, so that structure could be resolved);
for cases where the source registered little or no core flux and comparable lobe fluxes,
the measured second moments agreed well with the peak-to-peak sizes listed in Table 1.
However, in cases where an
appreciable core flux is detected and/or the lobe-to-lobe flux ratio differs substantially from unity,
the second moments tend to be systematically smaller than the peak-to-peak sizes,
confirming that the peak-to-peak distance is indeed a more robust measure of size---using 
the second moments would have the undesirable effect of introducing or
strengthening a power-size anti-correlation in the data.
For sources larger than $\sim 20^{\arcsec}$, the measured peak-to-peak sizes
agreed closely with those obtained using the model Gaussian fits, as expected.

$\;$ 4. In any flux-limited survey there is the possibility that large,
low surface brightness objects could be resolved out. The FIRST survey is 
sensitive to structure out to $\approx 100^{\arcsec}$, and it may be asked
whether larger double-lobed sources, whose lobes might equal or exceed this
angular size, might be missing from our sample.
Since the hot spots of FR-II objects are, by definition, high surface brightness features,
and the FIRST survey measures peak fluxes down to $\sim 1$ mJy, it is 
highly unlikely that these sources would be missed altogether; only very large, 
low surface brightness objects can remain undetected, and such sources would not correspond
to FR-IIs in our chosen redshift range. If such objects did exist, they would presumably,
due to their large sizes and integrated fluxes, be known radio quasars detected by previous
surveys, and should certainly be detected by the NVSS survey. Thus, to explore this
issue in a complete fashion, we individually compared the NVSS and FIRST radio maps for 
every previously known radio-active quasar in the Veron-Cetty \& Veron (1996)
catalog falling within the FIRST survey,
searching for double-lobed radio sources with sizes up to $1,000^{\arcsec}$
which might have been missed by FIRST. As expected, we find not a single instance of a large,
double-lobed quasar, with $z > 0.3$, detected by NVSS and not by FIRST, indicating
that there is no instrumental cutoff at the upper end of the observed angular sizes
in our sample.
This is confirmed by the fact that, within our chosen redshift range, 
the upper limit of our $\theta$-$z$ data agrees closely
with that of other samples drawn from less sensitive surveys, such as
3C and 4C (Hooley \etal 1978; Nilsson \etal 1993).

$\;$ 5. To confirm that our selected sample is not contaminated by FR-I sources,
we calculate the intrinsic 1.4 GHz power, 
\begin{equation}
P=4{\pi}S_{1.4}D_{A}^{2}(1+z)^
{3+\alpha},
\label{power}
\end{equation}
of each double-lobed 
quasar, where $S_{1.4}$ is the 1.4 GHz flux density and $\alpha$ is the 
radio spectral index ($S_\nu \propto \nu^{-\alpha}$). We assume standard values of
$\alpha=0.5$ for 
any core components, $\alpha=0.8$ for lobe
components (e.g., Gopal-Krishna \& Kulkarni 1992; Rector, Stocke, \& Ellingson 1995), 
and take $P$ to be the total core+lobe power.
For $h_0=0.5$ and $\Omega_0 =1.0$, we find a lower limit of $P = 1.45 
\times 10^{25}$ W Hz$^{-1}$ for our
sample, very near the observed break which separates FR-IIs from the 
lower-power FR-Is for this choice of cosmological parameters 
(Fanaroff \& Riley 1974, Neeser \etal 1995, Rector, Stocke, \&
Ellingson 1995). All but 9 sources have intrinsic powers in the range
$10^{26}$ W Hz$^{-1} < P < 10^{29}$ W Hz$^{-1}$, confirming that our sample is indeed
comprised of FR-IIs.

$\;$ 6. Another advantage of the flux sensitivity of the FIRST survey is that
our sample includes sources out to a redshift of $2.7$, significantly higher
than the redshifts at which the minima in the $\theta$-$z$ curves 
typically occur for different Friedmann models. This is in contrast to some
previous work, which used samples containing significant numbers of sources 
only to $z \la 1$ 
(Oort \etal 1987; Kapahi 1989), where roughly
Euclidean behavior is expected (see Figure 1).

$\;$ 7. Like all previous such samples, ours is a subset of the double-lobed 
radio sources which have measured redshifts, and may suffer from the
associated 
selection effects described in \S1. To examine this possibility, we performed 
a Kolmogorov-Smirnoff (K-S) 
test to compare the sizes of the double-lobed objects in our sample with 
those in the FIRST survey as a whole. However, since $\theta$ and $z$ are obviously 
correlated in our sample, with an expected upper limit to the correlation,
and we lack complete redshift information for sources in the FIRST survey, 
we must restrict the test to the range in $\theta$ ($\theta < 60^{\arcsec}$)
below which sources in our sample 
appear roughly uniformly at every redshift, in order to perform a fair comparison.
In the range $12^{\arcsec} < \theta < 60^{\arcsec}$ (the origin of the lower limit is discussed
below)
there are currently 13,664 objects in the FIRST survey
which are classified as true double-lobed
sources at the 95\% confidence level, based on morphology, positional and flux 
information (Buchalter \etal 1997) and two-point correlation
analysis (Cress \etal 1996). A K-S test shows that the null hypothesis---that 
our size-restricted subsample is drawn from this larger set---can only be rejected 
only at the 51\% level; i.e., it fails
to discriminate between the two distributions at the 1$\sigma$ level. 
In addition, since 47\% of the extended ($> 2^{\arcsec}$) sources in the FIRST survey
have measured fluxes below 3 mJy, but only 5 of our 103 sources have either lobe flux
in this range, we further restrict the K-S test to sources whose individual lobe
fluxes are $> 3$ mJy. In this case we find that the null hypothesis that our remaining
subsample is drawn from the 10,521 such sources in the survey can only be rejected only at 
the 30\% level.
Furthermore, it is 
estimated that the median redshift of the FIRST survey is $\sim 1.0$
(Cress \& Kamionkowski 1997) while the
median redshift of our sample is 0.98. This evidence, taken together,
indicates that the two populations have similar distributions,
and therefore that our optically selected 
sample (i.e., FR-II quasars with measured redshifts) 
is fairly representative of double-lobed
radio sources as a whole. This suggests that no serious selection effects
arise from measuring the radio sizes of a largely optically selected
subset of double-lobed quasars.

$\;$ 8. Since our sample excludes objects classified as radio galaxies, we 
avoid the possibility that different mean orientations between quasars
and radio galaxies, in the context of the unified scheme,
can be introducing non-cosmological effects into the
$\theta$-$z$ plane. However, the FR-II quasars in our sample do span a range
of core-to-lobe flux density ratios, $R$, suggesting that they are composed of both
CDQs and LDQs. If the redshift distributions of these two populations
are different,
such effects may still arise. For example, 
if the fraction of CDQs increases with redshift then the angular sizes
of our sources at higher redshifts would be depressed relative to
the cosmological predictions, since the CDQs are projected more closely to the line
of sight.
To investigate this issue, we perform a K-S test to compare the redshift
distributions of these two classes of quasars in our sample. Since the observed
median values of $R$ for these objects are $R_{CDQ} \sim 10$ and $R_{LDQ} \sim 0.1$
(Ubachukwu 1996),
we classify our sources using the geometric mean of $R=1$ for the critical value. 
This yields 12 CDQs and 91 LDQs in our sample, whose redshift distributions can be 
distinguished at only the 2\% confidence level; i.e., the probability that they are 
drawn from the same distribution is 98\%. Since many of the objects in our sample do 
not register a core component in the FIRST survey, we also perform a K-S test using
a critical value of $R=0.1$, which divides the sample more evenly into
47 `CDQs' and 56 `LDQs'. In this case,
the null hypothesis that the two redshift distributions are drawn
from different populations can only be rejected at the 38\% level (i.e., they cannot
be distinguished at the 1$\sigma$ level). Since CDQs and LDQs in our sample
do not exhibit significantly different redshift distributions, the expected differences
in their mean orientations should not alter the $\theta$-$z$ results.
More specifically, since 
\begin{equation}
R = \frac{R_T}{2} \left [ (1+\beta\cos\phi)^{-2} +
(1-\beta\cos\phi)^{-2} \right]
\label{Reqn}
\end{equation}
(Ubachukwu 1996) where $R_T$ is the value of $R$ when $\phi=90^{\circ}$
and $\beta=v/c$ is the flow speed in units of the speed of light (related
to the Lorentz factor), the 
apparent lack of a correlation
between $R$ and $z$ suggests that $\phi$ and $z$ are uncorrelated, and thus that
we are looking at a similar distribution of projection angles at every redshift.  

$\;$ 9. With a $5.4^{\arcsec}$ FWHM beam, the FIRST survey can detect extended structure
down to $2^{\arcsec}$ (White \etal 1997).
However, due to the survey resolution limit, uncertainties in the quasar
optical positions, and variations in the morphologies of double-lobed
objects, sources with $\theta \, \la 10^{\arcsec}$ could not be assigned an accurate
morphological classification. Thus, based on inspection of numerous FIRST radio maps,
we have introduced an effective cutoff in the data at
$12^{\arcsec}$, illustrated by the thin dashed
line in Figure 2. In terms
of defining a population of standard rods, it is, in fact, {\em
desirable} to have such a cutoff, in order to eliminate the possibility of 
including so-called core-jet, diffuse, cometary, and other 
types of extended radio sources which may be mistaken for double-lobed objects
at low resolution (cf. \S1). We now outline a self-consistent method
for incorporating this cutoff into the analysis.

\subsection{Optimizing the Analysis}

Consider the parameter space defined by the comoving intrinsic sizes, $l$, and
projection angles, $\phi$, of FR-II quasars, as illustrated in Figure 4. 
\placefigure{fig4}
In general, the intrinsic sizes will range up to some maximum value, $l_{max}$,
defined by the upper envelope to the observed angular sizes, and above which
there simply are no objects (see Figure 2). Determining the exact value of
$l_{max}$ from a given $\theta$-$z$ data set would require 
assumptions about the cosmology (see Eq.~(\ref{angular size})), 
but the actual value is irrelevant for the 
purposes of this discussion and no such assumptions need be made.
The projection angles will range between 0 and some upper limit $\phi_{u}$,
which for a randomly distributed population would correspond to
$90^{\circ}$, but in the context of the unified scheme would correspond to
a value of roughly $45^{\circ}$ (Barthel 1989; Lister \etal 1994). 
The accessible portion of the parameter
space is then defined by the largest bold rectangle in Figure 4. There will also in general
be some 
probability density along each axis, given by $P(l)$ and $P(\phi)$ (assumed
to be independent),
which will determine the forms of the distributions. For a distribution
of randomly oriented rods, it can be seen from simple geometric
arguments that $P(\phi) = \sin\phi$ (Harwit 1988);
$P(l)$ is still a matter of
theoretical and observational debate.

Since we are interested in defining a uniform population
of objects with true double-lobed structure, it makes sense to impose an
effective resolution cutoff at the angular scale for which
morphologies are well-determined. However, since a constant minimum
resolvable angular size does not translate into a constant minimum 
intrinsic linear size, the effect of this cutoff (or simply of the survey
resolution limit in general) will be to introduce a redshift-dependence
to the intrinsic size distribution. It is desirable therefore to impose
a minimum intrinsic size,
$l_{min}$, such that, for a given survey resolution, 
restricting a sample to sizes $l > l_{min}$ both avoids potential contamination by 
misclassified sources
{\em and} preserves the same range of observed sizes at every redshift. 
It may be expected that the average
size of this more homogeneous population (true double-lobed objects with
$l_{min} < l < l_{max}$)
is a more suitable measure of a 
standard rod than that of a distribution which includes objects with structure
down to the resolution limits of various surveys, probing different intrinsic
length scales, and possibly mixing different classes of sources. 

Since the $\theta$-$z$ relation always exhibits a minimum in Friedmann models,
when fitting a given cosmological model
to the data, one can define a subsample in which all objects have $l > l_{min}$
by aligning the minimum of the $\theta$-$z$ curve for that particular model with
the smallest observable {\em angular} size at which morphologies can be accurately determined,
and including only points above this curve.
The choice of $l_{min}$ is, then, determined by whatever value achieves this
alignment for the given model, though the actual value is immaterial.
The value of $l_{max}$ can be determined by finding 
the highest amplitude $\theta$-$z$ curve for the given
model which still passes through a data point in the sample and thus defines an upper 
envelope to the angular sizes. The sample so defined will include
maximally deprojected objects of intrinsic size $l_{min}$, and larger
objects viewed down to some projection angle given by the bold, dot-dashed curve
in Figure 4 (for example, objects with size $l_{max}$ can be seen projected
down to an angle of $\phi^{\prime}$). For the purposes of this analysis, however, one is free to 
examine only those sources with $l_{min} < l < l_{u}$, where $l_u$ can assume
any value between $l_{min}$ and $l_{max}$; i.e., the discussion presented here is valid
for any
choice of an upper envelope to the data which is lower than the $\theta$-$z$ curve
corresponding to $l_{max}$ and above that corresponding to $l_{min}$, for the assumed model. 
For an arbitrary choice of $l_u$, which we denote by $l_*$, the sample will include
maximally deprojected objects of intrinsic size $l_{min}$, objects with $l_{min} < l < l_*$
viewed from $\phi_u$ down to some projection angle given by the bold, dot-dashed curve, and 
objects with $l_{*} < l < l_{max}$ with projection angles between the dashed and dot-dashed
bold curves.
The objects in the sample will thus be located 
either in the combined area of regions
$B$ and $C$ (hereafter denoted by $BC$) in Figure 4 (for an upper envelope corresponding 
to $l_u=l_{max}$),
or simply in region
$B$ (for an upper envelope corresponding to $l_u = l_* < l_{max}$),
and the observed angular sizes, $\theta$, 
for a given Friedmann model correspond to the distribution of $l\sin\phi/D_A$ in these regions.
Due to
projection effects, some fraction of the objects with intrinsic sizes
larger than $l_{min}$ will be missed (corresponding to region $A$ 
the Figure).
If $P(l)$ and $P(\phi)$ were known, it would be a simple matter to
calculate the fraction of objects in region $A$ as well as 
the $l\sin\phi$ distribution in this region. 
However, if $P(l)$ and $P(\phi)$ are independent of redshift,
then for a given data set, the best-fit values of $\Omega_0$ and 
$\Omega_{\Lambda}$ for a particular model can be uniquely determined by the observed
$\theta$ distribution in any given region,
and are independent
of $P(l)$, $P(\phi)$, and the intrinsic size limits given by $l_{min}$, 
$l_{max}$, and $\phi_u$.
In other words, 
as long as the distribution of intrinsic projected sizes,
$l\sin\phi$, is not sensitive to redshift,
the actual values of $l_{min}$, $l_{max}$, and $\phi_u$ can only change the amplitude of 
the best-fit
curve to the $\theta$-$z$ data (cf. Eq.~(\ref{angular size})), not its shape. 
In contrast, a determination of
$H_0$ would require specific assumptions about these quantities and about $P(l)$.
Moreover, even if the $l\sin\phi$ distribution does
vary with redshift (e.g., because of some combination
of size evolution, a power-size correlation, or orientation effects), $\Omega_0$ and 
$\Omega_{\Lambda}$ can still be determined to the extent that this variation
can be modelled and corrected for in the data.

The question of whether or not $l\sin\phi$ is independent of redshift
is addressed quantitatively in the following section.
Note, however, that Figure 3 effectively
demonstrates that any relation between $l\sin\phi$ and $z$ cannot be strong; 
these graphs include all the observed data, without
incorporating any of the above size considerations, 
and are already seen to be fairly consistent with the conventional
curvature models, without invoking any redshift evolution of the apparent sizes.
We will hereafter use the term ``intrinsic size evolution'' to denote the case
where the intrinsic projected sizes, $s=l\sin\phi$ have a direct 
correlation with redshift (e.g., if $l \propto (1+z)^n$ with
$n \neq 0$, or if $\phi$ and $z$ are directly correlated), and 
``apparent size evolution'' to denote the case where $s$ and $z$ are
directly and/or indirectly correlated. Thus, apparent size evolution can arise from 
intrinsic size evolution, but also from other effects, such as an $l$-$P$ correlation 
coupled with a $P$-$z$ correlation. In general, various possible correlations may exist between
$l$, $\phi$, $P$, and $z$, and we investigate these in detail in \S3. 
However, only those which give rise to
apparent size evolution---an observed correlation between $s$ and
$z$---can affect the determination of  
cosmological parameters from a given data set. 

In the scenario we have presented, the best-fit cosmological parameters to a 
given data set can be found by 
exploring parameter space in the following fashion: 1) Assume a particular
cosmological model (we use
the word ``model'' as in \S1 to refer to the overall geometry and not the 
particular values chosen for $\Omega_0$ and $\Omega_{\Lambda}$). 2) Adopt
trial values for the relevant density parameters in that model. 3) Find the 
$\theta$-$z$ curve 
arising from these values (as in Figure 1)
and align the minimum value of this 
curve with the constant value of the effective angular size cutoff determined for
the survey in question.
Denote this curve by $\theta_{l}(z)$.
This then fixes the minimum intrinsic size for which accurately determined
morphologies in the sample are assured. In a similar manner, adjust the amplitude of the
trial $\theta$-$z$ curve so that it lies at some desired level above $\theta_{l}(z)$, 
and denote this curve by
$\theta_{u}(z)$. For example, to include all the data above $\theta_{l}(z)$, one would choose
$\theta_{u}(z)$ so that it defines the upper envelope to the observed angular sizes
(corresponding to $l_u = l_{max}$).
4) Beginning with all the
data above the effective cutoff, eliminate any data
in the $\theta$-$z$ plane with $\theta < \theta_{l}(z)$ or
$\theta > \theta_{u}(z)$
thus ensuring that the 
remaining sources all have intrinsic sizes between $l_{min}$ and $l_u$. 
This step assures that the intrinsic size limits
have been imposed in a self-consistent manner. 5) Using the remaining data, 
perform a $\chi^2$ 
goodness-of-fit test to determine the best-fit values of the remaining free
parameters in the test model (e.g., amplitude and size evolution parameter), 
and assess how well the output parameters for the
assumed model fit the resulting $\theta$-$z$ data. The entire procedure can then 
repeated so as to span the range of density parameters appropriate to the 
assumed model, as
well as to explore different models. Conducting the analysis in this 
fashion both defines a sample whose mean size is more akin to a standard rod
and accounts for the limited survey resolution in a self-consistent manner.

\section{Correlations Between Power, Size, and Redshift}

\subsection{Parametric Analysis}
	
Correlations among the properties of FR-II quasars 
have important implications for understanding the 
characteristics of the host active galactic nuclei and the evolution of the 
intergalactic medium, as well as for determining the best-fit cosmological
parameters from classical cosmological tests such as the $\theta$-$z$ relation.
These correlations must be considered in detail if these objects are to be used as probes
of the geometry of the universe.
Thus, before addressing cosmological issues, we explore
the relationships between the intrinsic properties of the sources in our sample,
by spanning the entire assumed range of cosmological parameter values
and testing for correlations among the intrinsic properties in each case.
For a given set of cosmological parameters, one can calculate respectively 
the intrinsic power and projected linear size,
\begin{equation}
 P=4{\pi}S_{1.4}D_{A}^{2}(1+z)^
{3+\alpha}; \qquad
s = l\sin\phi = {\theta}D_A,
\label{powerandsize}
\end{equation}
of each double-lobed 
quasar, where we again assume a spectral index of $\alpha=0.5$ for any
core components, $\alpha=0.8$ for lobe
components and take $P$ to be the total core+lobe power.

If we assume relationships between $P$, $l$, and $z$ of the form
\begin{eqnarray}
\label{prop1}
l \propto (1+z)^n \nl 
\label{prop2}
P \propto (1+z)^x \nl
\label{prop3} 
l \propto P^{\beta}
\end{eqnarray}
as in \S1, we can then determine the best-fit 
values of $n$, $x$, and $\beta$. In practice, it is straightforward to fit
for the $P$-$z$ and $l$-$P$ correlations, since these are expected to operate
independently of the third variable; a $P$-$z$ correlation should arise from the flux-limited nature
of the survey and an $l$-$P$ correlation should operate over the lifetime of the sources,
which is far less than the cosmological timescales spanned by the $z$ distribution.
Any observed $l$-$z$ correlation, however,
may be due to the separate correlations of $l$ and $z$ with $P$ 
and not to intrinsic size evolution. To address this possibility, 
previous authors have investigated the $l$-$z$ correlation for sources
within relatively narrow ranges of intrinsic power, so as to minimize the effect of any
dependence on $P$ (Hooley \etal 1978; Kapahi 1985; Singal 1988; 
Barthel \& Miley 1988; Singal 1993;
Nilsson \etal 1993). However, according to relations (\ref{prop1})-(\ref{prop3}),
the combined effects of intrinsic size evolution and a power-size 
correlation 
will result in an overall apparent $l$-$z$ correlation of the form 
$\theta \propto l \propto (1+z)^c$, where $c=\beta{x}+n$.
Thus, the value of $c$ for a given model follows directly from the data
and, together with the derived values of $\beta$ and $x$, one can
arrive at a value for $n$. 

We explore models 1 ($\Omega_0 \equiv 1$), 2, and 3, adopting values of $\Omega_0$ from 0.01 to 0.99
inclusive,
in intervals of 0.01, for models 2 and 3 ($h_0$ merely fixes the constants of proportionality
in relations (\ref{prop1})-(\ref{prop3}) and has no effect on $\beta$, $x$, or $n$, and all
3 models obviously yield the same results for $\Omega_0 = 1$), for a total of 199 possible scenarios.
For each scenario, we align the corresponding $\theta$-$z$ curve with the effective cutoff
at $12^{\arcsec}$ and include only data above this curve (denoted by $\theta_{l}(z)$), and below
some upper curve (denoted by $\theta_{u}(z)$) which may correspond to
the true upper envelope to the data, but can in general assume any lower amplitude still above that
of $\theta_{l}(z)$, as outlined above. 
For the remaining data, we compute, using five roughly equally populated bins,
the mean intrinsic projected sizes, $\VEV{s}$,
in bins of $P$, $\VEV{s}$ in bins of $z$, and 
$\VEV{P}$ in bins of $z$, together with the standard errors in these quantities,
and use a $\chi^2$ minimization
routine to determine $\beta$, $c$, and $x$, respectively.  
We bin the data since we do not have
{\em a priori} knowledge, independent of cosmological parameter values,
of the inherent scatters associated with the various
intrinsic properties of these sources, and by which the $\chi^2$ values must be weighted; 
a binned analysis allows us to obtain unbiased estimates for these scatters in each bin.
Note that
while the intrinsic sizes $l$ appear in relations (\ref{prop1}) and (\ref{prop3}), 
we can only obtain the intrinsic {\em projected} sizes, $s=l\sin\phi$.
Using the mean values, $\VEV{l\sin\phi}$, in each bin properly accounts for
the presence of projection effects only if the objects in each
bin have similar distributions of $\phi$. Otherwise, the derived values of
$\beta$, $c$, and $n$ could reflect variations of $P$ and $z$ with respect to 
$\phi$ as well as $l$, and some explicit $\phi$
dependence would need to be added to relations (\ref{prop1})-(\ref{prop3}) to break this degeneracy. 
Since we cannot directly test for $\phi$ correlations parametrically without
knowledge of the distributions of $R_T$ and $v/c$ in Eq.~(\ref{Reqn}), we explicitly
use $\VEV{s}$ rather than $\VEV{l}$, with the understanding that these are interchangeable
only if the $\phi$ distributions are consistent in each bin. However, we have already concluded, from
the K-S tests of the $R$ values in different redshift bins (\S2.2),
that $\phi$ does not vary significantly with redshift; we discuss further
the $\phi$ distribution below. 

The derived values
of $c$, $\beta$, and $x$, as well as the inferred values of $n$, together with the 1$\sigma$ errors
in each quantity,
are listed in the top-left portion of Table 2 for a representative 7 of the 199 scenarios. 
These results are obtained using all the data above $\theta_l(z)$
(i.e., ranging to the upper envelope defined by some value $l_u = l_{max}$) and are thus denoted by an upper 
limit of $l_{max}$ in the Table. All 1$\sigma$ errors quoted in this Section correspond to
the square roots of the appropriate diagonal elements of the covariance matrix for that 
$\chi^2$ fit (i.e., they are the errors obtained with $\Omega_0$ fixed at its trial value
and the corresponding constant of proportionality from relations (\ref{prop1})-(\ref{prop3}) fixed
at its best-fit value).
\placetable{tab2}
For all scenarios, we find a strong ($\ga $10$\sigma$) correlation between $\VEV{P}$ and $z$, which is expected
due to the flux-limited
nature of the survey. We also find an inverse correlation, at the 2--3$\sigma$
level between $\VEV{s}$ and $P$, which can be understood in terms of the
gradual fading of these sources as the lobes expand over 
timescales ($\sim 10^8$ yr)
much shorter than $D_A/c$, which is indeed the case for quasars in our sample. 
Previous studies have found $\beta \approx 0.3 \pm 0.1$ for double-lobed quasars
assuming an Einstein-de Sitter universe
(Oort \etal 1987; Kapahi 1989; Gopal-Krishna \& Kulkarni 1992;
Chy\.{z}y \& Zi\c{e}ba 1993; Nilsson \etal 1993; Singal 1993), which
is consistent with our result for model 1. Table 2 also shows that the observed data 
ranging up to $l_{max}$ exhibit mild apparent
size evolution at the $\sim$ 2$\sigma$ level, with all scenarios yielding $c \approx -0.8 \pm 0.4$.
Interestingly, however, the data imply little or no intrinsic size
evolution; all cases yield a slightly negative $\VEV{s}$-$z$ correlation
with $-0.4 < n < 0.0$, but are 
consistent with $n = 0$
well within the 1$\sigma$ level. The anti-correlation between intrinsic projected
size and redshift (given by $c$), seems to arise mainly from the separate correlations
of these quantities with intrinsic power. The lack of intrinsic size 
evolution in double-lobed quasar samples has been seen by other authors 
authors as well (Masson 1980; Singal 1993; Nilsson \etal 1993). 

There is also the possibility that the 
$\VEV{l\sin\phi}$-$P$ anti-correlation arises not from
intrinsically smaller objects having larger lobe powers, but 
rather from objects projected close to the line
of sight (CDQs) having relativistically boosted core power contributing significantly
to the total power.
In other words, the correlation could arise from different $\phi$ distributions in
different power bins, rather than a true power-size correlation. 
To explore this possibility, we re-solve for $\beta$, $x$, and $n$ using
only the lobe power, $P_l$, which, unlike the total power, is not expected to vary with
$\phi$. The resulting quantities, denoted by $\beta_l$, $x_l$, and $n_l$ ($c$ remains unchanged)
are listed in the upper-right portion of Table 2.
Though differing slightly, the values of $\beta_l$ are all within 1$\sigma$ of the 
corresponding values for $\beta$, and still yield an
inverse correlation between $\VEV{s}$ and $P_l$, at the 2--3$\sigma$ level in all cases, 
indicating that, while
orientation effects may be present, they are not primarily responsible for producing
the observed anti-correlation between intrinsic projected size and power.
Similarly, the values for $n_l$ differ slightly from those of $n$, in some cases
having a different sign, but still agree to well within the 
1$\sigma$ level,
implying that the $\VEV{l\sin\phi}$-$z$ correlation does not arise primarily from a $\phi$-$z$ correlation. 
As expected, the values of $x_l$ are not significantly different from $x$, since cosmological
surface brightness dimming affects $P$ and $P_l$ in a similar fashion. 

We have shown that the $\VEV{l\sin\phi}$-$P$ correlation is not the result of
orientation effects, but rather due to a negative correlation between $l$ and $P$,
and furthermore that this correlation, coupled with the $P$-$z$ relationship
appears largely to account for the apparent size evolution in the data.
Inspection of various scatter plots of the data for the different scenarios
supports these conclusions, suggesting
that the negative $\VEV{s}$-$P$ correlation arises mainly from the fact that the $\sim 20$ 
largest projected sources in the sample have low values of $P$ and lie
preferentially at lower redshifts.
We have already seen, however, that
one need not choose the amplitude of $\theta_{u}(z)$ to 
trace the upper envelope to the $\theta$-$z$ data (i.e., to $l_{max}$), 
but can, in principle, choose any value between this and $\theta_{l}(z)$ without loss of generality
or the introduction of sample bias.
Thus, in light of the above conclusions, we
re-derive the values of $c$, $\beta$, $x$, and $n$, fixing the $\theta_{u}(z)$ curve
in each scenario to have a minimum at $65^{\arcsec}$ (corresponding to some intrinsic size $l_u = l_*$),
which eliminates roughly 20 of the largest sources in each scenario.
The removal of these sources reduces the seemingly anomalously high values
in the first redshift bins of Figures 3a--d, and also eliminates the few exceptionally
large sources at $z > 1$ (see Figure 2).
The results obtained from the remaining samples for the same 7 scenarios
are presented in the lower portion of Table 2, where they are denoted by an upper limit of $l_*$.
We indeed find that, in all trial cases, the magnitude of the $\VEV{s}$-$P$ correlation diminishes significantly,
with $-0.11 < \beta < 0.0$ in all cases, and is generally 
consistent with zero at the $\la 2\sigma$ level.
The $\VEV{P}$-$z$ correlation remains highly significant, as expected.
The derived values for $n$ become positive, rather than negative, in all cases, but 
are generally much smaller in magnitude than those obtained using an upper limit of $l_{max}$
and are certainly consistent with $n=0$ well within the 1$\sigma$ range. Most importantly,
we find, as expected for the $l_*$ samples, that there is no significant apparent
size evolution, with $c$ consistent with zero well within the 1$\sigma$ level for
all scenarios. These results confirm that large, fainter sources (which can be seen only at lower redshifts)
were largely responsible for the power-size correlation and resulting apparent size evolution
observed previously, and suggest that we can define a sample whose $\theta$-$z$ variation
should primarily be due to cosmological effects alone. 
Unlike the case for the samples obtained using an upper limit corresponding to $l_{max}$,
the values of $\beta_l$, $x_l$, and $n_l$ obtained using $l_*$, are virtually identical to those for
$\beta$, $x$, and $n$, indicating that for these samples, orientation effects play no role
in the $\VEV{s}$-$z$ and $\VEV{s}$-$P$ correlations.

\subsection{Non-Parametric Analysis}

The analysis of \S3.1 assumes specific functional forms for the relationships
between $P$, $l$, and $z$, and is valid only
insofar as these parameterizations accurately reflect the 
underlying physics. If this is not the case,
the resulting fitted values are merely artifacts
of the model, not parameters truly descriptive of the data. 
For example, the relation $P \propto (1+z)^{x}$, though a good
approximation, does not properly account for the implicit dependence
of $D_A$ on $z$ (see Eqs. (\ref{angular size}) and (\ref{powerandsize})).
Thus, a better approach to searching for
correlations in the data is to employ non-parametric tests that are independent of an assumed
functional form. 
In addition, non-parametric statistics offer another advantage in that they 
are readily applied to unbinned distributions and thus incorporate information
that is lost when the distributions are binned. In particular, the Spearman rank
correlation coefficient, $r_{ab}$, tests, in a non-parametric fashion,
the degree to which the quantities
$a$ and $b$ are correlated in a given data set, varying from -1 (for strong negative 
correlation) to +1 (for strong positive correlation), with 0
indicating no correlation. The Spearman partial-rank statistic,
$r_{ab,c}=(r_{ab}-r_{ac}r_{bc})/\sqrt{(1-r_{ac}^2)(1-r_{bc}^2)}$, 
has the same range, and tests whether there is a 
significant correlation between
$a$ and $b$ which does not arise from both being separately
correlated with a third quantity $c$; i.e., it effectively
tests for a correlation between $a$ and $b$ if $c$ is held constant.
For data sets with $\geq 30$ data points, the distribution of $\sqrt{(n-1)}r_s$, where
$r_s$ is a Spearman statistic, is well
approximated by a normal distribution with unit variance (Conover 1980). 
Thus, for a given data set with an observed Spearman statistic $r_{obs}$, 
one can easily compute the (two-sided) probability $p$, 
that a random, uncorrelated data set, with $r_{ran}$, could exhibit this 
degree of correlation (positive or negative) or higher (i.e., the
probability that $\abso{r_{ran}} \:\geq \:\abso{r_{obs}}$),
and thus obtain the significance of the result.
For the correlation coefficient, $r_{ab}$, this is simply the probability
of seeing $r_{ab}$ occur by chance if there is no intrinsic correlation 
between $a$ and $b$. 
For the partial rank statistic, $r_{ab,c}$, it is the probability of seeing
$r_{ab,c}$ occur by chance
if there is no correlation between $a$ and $b$ other than that caused by 
their being separately correlated to $c$. 
	
The upper portion of Table 3 shows the results of the rank analysis, listing the Spearman 
statistics for the various combinations of $z$, $s$, and $P$, 
and the corresponding values of $p$ (given in parentheses),
for the same 7 of the 199 trial scenarios from \S3.1 using an upper limit corresponding to $l_{max}$. 
\placetable{tab3}
In all cases investigated, we find evidence for a negative correlation between $s$ and $z$
at the 90-95\% confidence level (given by $1-p$), a significant inverse correlation between
$s$ and $P$, near the 3$\sigma$ level (i.e., $1-p > 99$\%) and a highly significant $P$-$z$
correlation. Moreover, the partial rank correlation coefficient, $r_{sz,P}$ indicates that
the $s$-$z$ correlation arises entirely from the $s$-$P$ and $P$-$z$ correlations,
so that $s$ and $z$ are intrinsically uncorrelated, with $p > 0.36$ in all cases.
Intrinsic size evolution is consistent with zero well within 1$\sigma$ for all cases and
does not account for the mild degree of apparent size evolution.
These results all agree closely with our results
from the parametric analysis. 

We also tested $s$ separately against
$P_l$ to see whether the $s$-$P$ correlation was truly due to a
power-size correlation (to which $P_l$ is sensitive), and not due to beaming effects
(to which $P_l$ is not sensitive). We indeed find that values of $r_{sP_l}$ are close to those
for $r_{sP}$, with the significance level remaining near or above 98\% in all cases
We do not directly test for a correlation with $P_c$, 
since not all sources registered a FIRST core component, but
we do eliminate CDQs, using both criteria of $R > 1.0$ and also $R > 0.5$, re-solve for $r_{sz}$,
and in both cases find similar values to those in Table 3, indicating again that 
a true
power-size correlation, and not beaming effects,
are responsible for producing the observed apparent size evolution.
Also, since some sources listed as having $P_c=0$ may in truth have some core flux which
was not separately represented in the FIRST catalog, we re-perform the analysis
including only those sources with 
$P_c \neq 0$ and again find virtually identical results to those in Table 3.

In the lower portion of Table 3 are the results obtained using
an upper limit corresponding to $l_u = l_*$ as defined above. Again we confirm the results of the
parametric tests, finding that while the $P$-$z$ correlation remains highly
significant, the $s$-$P$ correlation is reduced in magnitude, with significance
typically below the 2$\sigma$ level in the various scenarios, and the $s$-$z$ correlation 
effectively vanishes in all cases,
consistent with zero apparent size evolution.
As expected, these
results agree with the parametric analysis, but are more robust in the sense
that they are independent of the assumed model governing the characteristics 
of the sources.  

The non-parametric analysis also allows us to probe the $\phi$ distributions more
directly and examine issues related to unification schemes for radio-loud
AGNs. Since $R$ is expected to be correlated with $\phi$ via Eq.~(\ref{Reqn}), we can, 
unlike in the parametric
case, test for correlations between some quantity $q$ and $\phi$ through $r_{Rq}$
without invoking assumptions about the distributions of $R_T$ or $v/c$.
We have already seen that the $\phi$ distribution is not sensitive to redshift
and that it does not account for the observed $s$-$P$ correlation.
However, if the unified scheme is correct, in the sense that sources
projected near the line of sight have relativistically boosted core fluxes,
then there should be a negative correlation between $R$ and intrinsic 
projected size, $s$, and, obviously, a
positive correlation between $R$ and the intrinsic core power, $P_c$,
since $R \propto P_{c}/P_{l}$.
Some sources in the sample may have registered $R=0$ not because they truly lacked
a significant core, but rather, in the case of smaller sources,
because the FIRST fitting algorithm did not assign
that source a core component. Assigning
the flux in such sources to the lobes had no significant effect
on the results above, as seen when we omitted sources lacking a measured
core component, but could seriously affect apparent correlations with $R$.
Thus we include here only those sources with $R \neq 0$.
Table 4 shows the results of the rank tests between various quantities and $R$
for these sources, with the upper and lower sections again corresponding to limits
of $l_{max}$ and $l_*$, for the same 7 of the 199 scenarios. For all cases, we 
indeed find a significant
($> 99.7$\%) inverse correlation between $R$ and $s$, as predicted by the unified model.
Moreover, while $R$ and the total power $P$ show no statistically significant 
relationship, $R$ and $P_c$ exhibit a positive correlation with high significance
($p < 10^{-4}$) in all cases, as would be expected. 
Though only suggestive, these results
indicate that the behavior of these radio-loud quasars
is consistent with the expectations of unification schemes.
\placetable{tab4}
	
\section{Cosmological Parameters}

Having explored the intrinsic properties of the sources
over the assumed range of cosmological density parameter values,
we now turn to a discussion of the best-fit cosmological results.
The cosmological models we consider are models 1, 2, and 3 from \S1, along with
a Euclidean model for comparison. Since the results of the non-parametric
analysis corroborate those of the parametric analysis, we allow for apparent 
size evolution in the data of the form $l \propto (1+z)^c$ so that
$l\sin\phi = (\l\sin\phi)_{0} (1+z)^c$ where a zero subscript denotes the
present-day ($z=0$) value. 
To determine the best-fit
values for the free parameters in each model,
we follow the prescription in \S2.3 and minimize the quantity
\begin{equation}
\chi^2=\sum_{i=1}^{N} \left( \frac {(\VEV{\theta_{p}}(a,c,\Omega_0;z_i)-
\theta_i)^2} {\sigma_i^2 + \sigma_{\theta}^2} \right); \qquad 
\VEV{\theta_{p}} = \frac {a(1+z)^c} {f(\Omega_0;z)},
\label{chi1}
\end{equation}
where $\VEV{\theta_p}$ is the mean angular size predicted by the model at $z_i$,
given $a$, $c$, and $\Omega_0$, 
the $\sigma_i$ are the errors associated
with the $N$ individual measurements $\theta_i$, $a=h_{0}\VEV{l\sin\phi}_0$ 
(measured in Mpc throughout) fixes the 
overall amplitude of the $\theta$-$z$ curve, and $f(\Omega_0;z) = h_{0}D_A$ is given by 
Eq.~(\ref{angular size}).
The quantity $\sigma_{\theta}=\sigma_{l\sin\phi}/D_A$ is the observed root
variance in the distribution of $\theta$, which in general
arises from the spread in the intrinsic projected
sizes, given by $\sigma_{l\sin\phi}$, as well as from curvature effects.
Without knowledge of $P(l)$, $l_{min}$, $l_{max}$, and  $\phi_u$, we cannot 
make an {\em a priori} determination of $\sigma_{l\sin\phi}$, 
which fixes the scatter in $\theta$ at a given $D_A$. 
Therefore, one must resort to a binned analysis to estimate $\sigma_{\theta}$ 
from the scatter in $\theta$ in different bins.
Furthermore, it is clear that the $\sigma_i$ (typically $\approx 1^{\arcsec}$) are much
smaller than the scatter in $\theta$ at any given redshift, and thus that $\sigma_i^2 \ll 
\sigma_{\theta}^2$, so that we can ignore the $\sigma_{i}$.
We thus seek to minimize
\begin{equation}
\chi^2=\sum_{j=1}^{M} \left( \frac {(\VEV{\theta_{p}}(a,c,\Omega_0;z_j)-
\VEV{\theta_j})^2} {\sigma_{\VEV{\theta_j}}^2} \right),
\label{chi2}
\end{equation}
where now the $\VEV{\theta_p}$ are the predicted mean angular sizes in $M$ bins centered
at $z_j$ with observed mean sizes $\VEV{\theta_j}$ and corresponding standard errors
$\sigma_{\VEV{\theta_j}}$. 

Using five roughly equally populated bins, we follow the method outlined in \S2.3 and
calculate $\chi^2$ with respect to the free parameters in each model.
Model 1, with a fixed value of $\Omega_0$, has only
two free parameters, $a=h_{0}\VEV{l\sin\phi}_0$ (with $a>0$) and $c$. 
The quantity $\VEV{l\sin\phi}$ appears
because it is, by definition, the mean value of $\theta$ around which
$\chi^2$ will be minimized (cf. Eq.~(\ref{angular size})).
Models 2 and 3 each have three free parameters,
$a$, $c$, and $\Omega_0$ (with $0 < \Omega_0 \leq 1$). 
The Euclidean model simply has 1 free
parameter corresponding to the amplitude.
For each trial value of $\Omega_0$ in the Friedmann models ($\Omega_0 \equiv 1$ in model 1,
and trial values in intervals of 0.01 from 0 to 1 for models 2 and 3),
we use the resulting $\theta_l(z)$ and $\theta_u(z)$
curves to ensure a uniform range of intrinsic sizes, and then determine the best-fit
values of $a$ and $c$ for the remaining data, as well as the value of
$\chi^2$ for this set of parameters. In practice, aligning the minimum intrinsic size 
cutoff of the
various Friedmann models with the survey resolution limit, 
which is vital in terms of producing a 
self-consistent result,
removes 3 to 17 data points depending on the trial value of $\Omega_0$, but typically
fewer than 5 for $\Omega_0 > 0.3$. 
This prescription is meaningless for the Euclidean case, which does not exhibit a 
minimum in the angular size; for this case we simply follow the approach of past
workers and use all the data above the cutoff at $12^{\arcsec}$.

The results of our 
analysis for each of the four models, using $M=5$ bins
containing the various $N$ data points between $\theta_l$ and
$\theta_u$ given by $l_u = l_{max}$, 
are shown in the upper portion Table 5, which 
lists the values of $N$ and $\chi^{2}$ for each best fit, together with the number of 
degrees of freedom, $\nu$, 
in the model, and
resulting significance level, $1-p$, as well as the best-fit values of the free parameters, 
and the $1\sigma$ confidence limits on these values. 
It should be noted that the values
of $p$ are computed under the assumption of normally distributed data.
Though the unbinned intrinsic projected sizes, $l\sin\phi$, and thus
the angular sizes, $\theta$, given by Eq.~(\ref{angular size}), are not expected
to follow a normal distribution, or even to be symmetric about their mean values
(see Figure 2), the central
limit theorem ensures that, for binned data with
a sufficiently large number of points, the distribution
of the mean value in each bin (which is, in fact, our dependent variable)
will be close to a Gaussian, independent of the underlying distribution.
Insofar as we have $\la 20$ points in each bin, the probabilities derived from our $\chi^2$
values might be slightly in error due to any residual non-Gaussianity.
Moreover, the lower (and for cases with $l_u = l_*$, upper) tails of the observed $\theta$
distribution have been removed by the cuts in our analysis method, thereby
enhancing the non-Gaussianity. 
Thus, while it is 
straightforward to compute the mean values and calculate the value of $\chi^2$
for each trial model, the formal significance of the result cannot be obtained in
a simple, analytic fashion (nor can it be computed numerically without knowing 
the intrinsic size distribution);
the listed values of $1-p$ for the various models 
are intended to be qualitatively illustrative of the relative
significance levels, and not rigorously accurate. 
Since we have no {\em a priori} knowledge of the actual values of the free parameters,
all parameter errors quoted here do {\em not} correspond to the diagonal elements of
the covariance matrix of the fit (i.e., to the error obtained with the other parameters
held fixed at their best-fit values), but rather to the much larger error range subtended by the
joint variation in all free parameters, given conservatively by the various 1-dimensional projections
of the 1$\sigma$ confidence region in the parameter space. 
Figure 5 provides
a graphical representation of our results.
\placetable{tab5}
\placefigure{fig5}

It is clear that the observed data are entirely consistent with Friedmann
models with reasonable values of $\Omega_0$. 
The underdense models 2 and 3 both yield $\Omega_0 \approx 0.35$
with a 1$\sigma$ range including values from $ \sim 0.25$ to 1.00, 
and exhibit a fairly flat $\chi^2$ surface in this range
of parameter space, so that they are truly
consistent with the value $\Omega_0=1$ required by model 1
\footnote[4]{Since $\Omega_0$ in models 2 and 3 was constrained to lie between 0 and 1, 
the confidence limits explored were similarly restricted to this interval.
Fits to closed Friedmann models, with $\VEV{\theta_{p}}$ calculated using $\Sigma(x) = \sin{x}$ in 
Eq.~(\ref{angular size}), invariably yielded values of $\chi^2$ significantly larger than the minimum value in
corresponding non-closed model. Thus, while values of $\Omega_0 > 1$ in these models did fall within
the 1$\sigma$ range of the best-fit value, we do not consider the results of closed models 
in the present treatment.}.
The constraints implied by model 2 on the energy density associated with the
cosmological constant are $\Omega_{\Lambda} = 0.62$ with 1$\sigma$ limits ranging
from 0 to 0.72.
All three Friedmann models are seen to yield roughly equal values of $1-p$, indicating,
as expected, that the present data 
cannot effectively discriminate between the various Friedmann models,
although interesting
constraints on the free parameters within a given model are obtained
\footnote[5]{Kellerman (1993), looking at the sizes of {\em compact} 
sources on milliarcsecond scales, found the
$\theta$-$z$ relation to be consistent with an Einstein-de Sitter universe, but
did not consider other possible models (Krauss \& Schramm 1993).}.
It should be pointed out, however, that model 1, which is merely a particular
case of models 2 and 3, yields a comparable value of $1-p$ only because
$\Omega_0$ is assumed to be known {\em a priori}, thus allowing for an additional degree of freedom
in the fit. Any presumed value of $\Omega_0$ within the 1$\sigma$ range of the best-fit
values would similarly yield a fit on par with that of models 2 and 3. 
To the extent that $\Omega_0$ is not
pre-determined, there is no particular significance to the results of model 1;   
it is included primarily because it is the canonical standard among current theoretical models.
Note that the best-fit Euclidean model (which uses all the data points) is the only one which
yields a reduced chi-square value, $\chi^2/\nu$, greater than unity
and is actually a comparatively poor fit to the data.   
For comparison, we also performed a semi-unbinned analysis, using the individual values
$\theta_i$ as in Eq.~(\ref{chi1}), rather the mean of the binned values, but
still assign each source a $\sigma_{\theta}$
corresponding to the standard deviation of the angular sizes
in the bin corresponding to that source (i.e., we use the binned values of $\sigma_{\theta}$,
but not of $\theta_i$), and obtain roughly identical results for the best-fit
parameters in each model.

As expected from the results of Section 3, the data
appear to require mild apparent size evolution with $c \approx -0.8$
for model 1 and $c \approx -0.6$ for both models 2 and 3.
We have seen that this trend arises
primarily from a power-size correlation, rather than 
from intrinsic size evolution or orientation effects between
CDQs and LDQs (orientation differences between radio galaxies and quasars
are ruled out since we have included only the latter in our sample).
The values obtained here for $c$ agree closely with 
the corresponding results from Section 3, but differ slightly because
we have here assigned each source the value of $D_A$ corresponding to its bin,
so that $\VEV{\theta_j}D_{A_{j}} \propto (1+z)^c$ over the $j$ bins, rather than
taking $\VEV{\theta_{i}D_{A_{i}}}$ for each of the $i$ sources, and then binning
the projected sizes, as in Section 3. The error ranges also differ since,
as described above, we have
here taken the errors to arise from the joint variation of all free parameters.  
An inspection of the variation of $\chi^2$ with respect to the free
parameters reveals that $\chi^2$ in a given model 
is significantly more sensitive to changes in $c$ than
in $\Omega_0$, i.e., $\abso{d\chi^2/dc} > \abso{d\chi^2/d\Omega_0}$. Therefore,
since the effect of $c<0$ is to decrease the apparent sizes of sources
at higher redshifts, mimicing a decrease in $\Omega_0$,
we may infer that, for models 2 and 3, the best-fit values for $\Omega_0$ obtained
using an upper limit corresponding to $l_{max}$ are likely
lower limits to the actual values. We have already seen, however, that we can define a 
sample for which apparent size evolution is minimal, and for which
the derived values of $\Omega_0$ should therefore
correspond more closely to the actual values. 
The lower portion of Table 5 shows the results obtained using an
upper limit corresponding to $l_u=l_*$. In this case, we see that all
models yield $c \approx -0.2 \pm 0.5$, and do indeed find
higher best-fit values of 0.84 and 0.93, respectively, for $\Omega_0$ in models 2 and 3,
appearing to favor a flat universe. The lower amplitude of the $\theta_u(z)$ curve,
however, removes $\sim 20$\% of the data, and the resulting sample yields
larger errors on $\Omega_0$ in models 2 and 3, effectively spanning the range from 0 to 1.
In this case, model 2 yields $\Omega_{\Lambda} = 0.16$ with 1$\sigma$ limits of 0.0 and 0.97.
We also find that while all three models again fit the data with high significance,
that model 1, with $\Omega_0=1$ appears to be slightly favored, subject to the 
qualification discussed above.
Since all three models yield similar values for $\Omega_0$,
as well as $a$ and $c$, in the $l_u = l_*$ case,
they each pick out the same 83 data points, 
and are thus all plotted on the same graph in Figure 6, where they
are seen to virtually overlap. 
\placefigure{fig6}

In principle, we could test
the robustness of the zero apparent size evolution feature of the $l_u=l_*$ data
and of our analytic methods, by raising the amplitude of the $\theta_l(z)$
curve, effectively mimicing a survey with poorer angular resolution.
If apparent size evolution was not truly absent, then performing our analysis with a higher 
survey cutoff would selectively
eliminate different fractions of sources in different redshift bins, changing
the value of $\VEV{l\sin\phi}$ in each bin by different amounts and thus producing
a different value for $\Omega_0$. On the other hand, if the angular size distributions
truly are redshift-independent, raising the resolution cutoff 
in our analysis would remove 
the same fraction of sources in each bin (those
with sizes below the new value of $l_{min}$), changing the amplitude of the
best-fit curve, but not its shape. We would thus expect to find a different,
higher value for $a$, but the same value for $\Omega_0$.
In practice, we cannot meaningfully conduct this test with the current data set, since
it would further remove data from the $l_u=l_*$ sample, which already yields
formal errors on $\Omega_0$ that span the allowed range, but do employ a similar
technique below.
	
So far we have only discussed measurements of $\Omega_0$ and $c$. Each curve,
however, is also parameterized by an amplitude $a=h_{0}\VEV{l\sin\phi}_0$.
If we assume functional forms for $P(l)$ and $P(\phi)$, as well as values
for $l_{min}$, $l_{*}$, and $\phi_u$, we can compute 
the theoretical value of $\VEV{l\sin\phi}_0$ in regions $B$ and $BC$ of Figure 3,
given respectively by
\begin{equation}
\VEV{l\sin\phi}_0 \mbox{ in $B$} =  \frac {\int_{\phi^{\prime} }^{\phi^{\arcsec}} \left[
\int_{l_1}^{l_{max}}P(l) (l \sin \phi) dl \right]
P(\phi) d \phi +
\int_{\phi^{\arcsec} }^{\phi_u} \left[
\int_{l_1}^{l_2}P(l) (l \sin \phi) dl \right]
P(\phi) d \phi
}
{\int_{\phi^{\prime}}^{\phi^{\arcsec}} \left[
\int_{l_1}^{l_{max}}P(l) dl \right]
P(\phi)d \phi +
\int_{\phi^{\arcsec}}^{\phi_u} \left[
\int_{l_1}^{l_2}P(l) dl \right]
P(\phi)d \phi
} 
\label{lsinphiB}
\end{equation}
\begin{equation}
\VEV{l\sin\phi}_{0} \mbox{ in $BC$} =  \frac {\int_{\phi^{\prime} }^{\phi_u} \left[
\int_{l_1}^{l_{max}}P(l) (l \sin \phi) dl \right]
P(\phi) d \phi }
{\int_{\phi^{\prime}}^{\phi_u} \left[
\int_{l_1}^{l_{max}}P(l) dl \right]
P(\phi)d \phi}, 
\label{lsinphiBC}
\end{equation}
where $l_1 = l_{min}\sin\phi_{u}/\sin\phi$, $l_2 = l_{*}\sin\phi_{u}/\sin\phi$,
$\phi^{\prime}=\arcsin({l_{min}\sin\phi_u}/l_{max})$ is the minimum angle
to the line of sight at which objects with intrinsic size $l_{max}$ can be seen
by the survey, and similarly, $\phi^{\arcsec}=\arcsin({l_{*}\sin\phi_u}/l_{max})$.
The denominators in Eqs. (\ref{lsinphiB}) and (\ref{lsinphiBC}) assure proper normalization.
The best-fit values of $a=h_0\VEV{l\sin\phi}_0$
for $l_u=l_{max}$ and $l_u=l_*$, can thus be compared
with the theoretical value of $\VEV{l\sin\phi}_0$ in regions $BC$ and $B$, respectively,
to arrive at values for $h_0$ in each Friedmann model. 
Though not a valid determination of $H_0$,
this does offer a consistency check on our analysis, in the sense
that we expect reasonable input assumptions to yield plausible values for $H_0$.

If the lobe sizes grow as $l=vt$
with some expansion velocity $v$, and then fade, with an overall lifetime of 
about $10^7$
to $10^8$ yr (Nilsson \etal 1993; Neeser \etal 1995; Gopal-Krishna \etal 1996),
a population of such objects observed
over cosmological timescales $\gg 10^8$ yr will yield an observed distribution of 
sizes roughly constant between $l_{min}$ and $l_{max}$. In this case,
the normalized probability density between $l_{min}$ and $l_{max}$
is given by $P(l)=1/(l_{max}-l_{min})$. For a spherically symmetric distribution of
randomly oriented rods, $P(\phi)=\sin\phi$ for $0<\phi<\phi_u$ (Harwit 1988). 
If quasars are viewed at arbitrary projection angles then $\phi_u=90^{\circ}$. 
In the unified model, however, the jet axes of quasars tend to lie
nearer to the line of sight, so that $\phi_u \approx 45^{\circ}$. We consider 
both values for $\phi_u$. 
The theoretical value of $\VEV{l\sin\phi}_0$ in region $B$ also depends 
on $l_{min}$, $l_{max}$, and $l_*$, while that in region $BC$
depends only on $l_{min}$ and $l_{max}$. 
The ratios between these quantities, however, are determined by the data;
points lying near $\theta_l(z)$ for a given model (from step 3 in our
analysis method) were assumed to correspond to the
maximally deprojected minimum intrinsic size, $l_{min}\sin\phi_u/D_A$, while
the uppermost points in the $\theta$-$z$ plane are taken to correspond to
$l_{max}\sin\phi_u/D_A$, so that the values of $l_{min}/l_{max}$ are uniquely 
determined
from the output data sets used to fit our Friedmann models. The ratio
$l_{min}/l_*$ is also fixed from our requirement that the minimum in the
$\theta$-$z$ curve arising from $l_*$ be fixed at $65^{\arcsec}$. 
To complete the theoretical calculation, it remains only to assume a value for 
$l_{max}$, the present-day maximum intrinsic linear size.
The largest known double-lobed radio sources have estimated
intrinsic linear sizes of order 1 Mpc (Schoenmakers 1997). 
We thus take $l_{max} = 1.2$ Mpc, which yields values of $l_{min} \approx 70$ kpc 
for the various Friedmann models.

Our inferred results for $H_0$ under these assumptions, 
using all the observed data
above the $12^{\arcsec}$ survey limit, are shown in the top portion of Table 6, along
with the 1$\sigma$ error ranges. The
derived values for $H_0$ vary simply as $1/l_{max}$. 
Though the error bars are considerable, 
the plausible assumptions we have made yield results for $H_0$ that agree
generally with the range spanned by current measurements (Schechter \etal 1996;
Giovanelli \etal 1996; Sandage \& Tammann 1996; Kim \etal 1997; Falco \etal 1997;
Holzapfel \etal 1997), with the unified model giving higher values.
This in turn suggests 
that our model and input assumptions are in fact reasonable.
It is also interesting to examine the derived values for $H_0$ using an 
upper limit to the data corresponding to $l_*$. Though the best-fit values
of $\Omega_0$ and $c$ are different in this case, and the predicted theoretical value of 
$\VEV{l\sin\phi}_0$ in region $B$ differs from that in region $BC$, 
the best-fit value of $a$ in each model, as determined by 
the $l_*$ data should compensate so as
to yield values for $H_0$ in agreement with the $l_{max}$ results, 
if the input assumptions are valid. 
The lower portion of Table 6 shows that while the $l_*$
values are systematically lower, they are consistent to within the
1$\sigma$ errors, with those obtained using $l_{max}$. In particular, the range
of values (in km s$^{-1}$ Mpc$^{-1}$) for which the 1$\sigma$ limits from identical models
using limits of $l_{max}$ and $l_*$ overlap are 59-77, 64-114, and 64-93,
respectively, for models 1, 2, and 3, assuming $\phi_u=45^{\circ}$, and
37-54, 41-81, and 40-66, assuming $\phi_u=90^{\circ}$. 
\placetable{tab6}

\section{CONCLUSION}

Using the FIRST radio survey and available redshift information
we have constructed a carefully defined set of double-lobed
quasars whose observed $\theta$-$z$ relation, unlike those
of many previous studies, appears to show evidence for
curvature. We attribute this result to the precise sample definition,
to the increased depth and sensitivity of the survey data, and to our self-consistent
method of analysis, which addresses many of the problems associated with previous work
in this area. We have explored the correlations between the intrinsic properties
of these sources and find evidence, regardless of cosmological parameter values, 
for apparent size evolution arising from an inverse power-size
correlation, and evidence against intrinsic size evolution,
both of which agree with
the results of some previous authors. We find that while the present data
can place interesting constraints on $\Omega_0$ within a given 
cosmological model, in particular
suggesting, for models with $\Omega_0 \leq 1$, values in the range from 0.25 to 1.0 inclusive, with 
some evidence favoring values of (or near) unity, 
they cannot distinguish between various models with reasonable significance.

A larger data sample (e.g., from additional redshift information on the thousands
of radio doubles in the FIRST survey), however, would
place stronger constraints on the parameters within each model, 
and may be able to distinguish among models. To investigate this, we have
performed a Monte-Carlo simulation
using $P(\phi) = \sin\phi$, $P(l) = 1/(l_{max}-l_{min})$, $l_{max} = 1$ Mpc, 
$\phi_u = 45^{\circ}$, $H_0 = 50$ km s$^{-1}$ Mpc$^{-1}$, and an effective resolution 
cutoff at $12^{\arcsec}$,
to generate mock $\theta$-$z$ data for double-lobed sources
assuming different cosmological models and choices of $\Omega_0$. We find that,
if apparent size evolution is negligible, a data sample with $\sim 500$ points
can recover the input value of $\Omega_0$ to within $\pm 0.2$, but, in the case
of underdense models, still cannot effectively distinguish between models with and
without a cosmological constant (see Figure 1). If apparent size evolution with $c = -1$ is
included, at least twice as much data is required to achieve comparable results,
due to the sensitivity of $\chi^2$ to $c$. 

Our sample, like all other $\theta$-$z$ studies to date, consists of double-lobed
sources whose sizes are measured in the radio, but whose redshifts were 
typically obtained in an optically selected fashion. Though we offer evidence, in
\S2, as to why no serious selection effects are believed to be introduced
by this mixing of optical and radio properties, a more desirable approach,
in principle, would be to obtain redshift information for a complete and homogeneous
sample of radio-selected double-lobed sources. 
One can further refine the sample by including only radio sources with 
symmetric and colinear triple structure (i.e., core + 2 lobes), thereby
minimizing asymmetrical effects which might distort the apparent angular size, such as
relative motion with respect to the IGM, and simplifying the problem of optical
identification,
since the positions of the central engines are well-determined {\em a priori}. 
We have selected a sample of such objects from the
FIRST database (Buchalter \etal 1997) and matched these
with the Automatic Plate Machine scans of the POSS plates (Irwin \& McMahon 1992; 
Irwin, Maddox, \& McMahon 1994) to produce a subset of radio 
triples having optical counterparts to the central 
source. This sample
constitutes a set of several hundred radio-selected double-lobed sources
complete to roughly $V = 20$. About 5\% of these objects
have been previously identified as radio galaxies or quasars
and fewer than 1\% of these sources
have known redshifts (NED), though many are expected to be substantially 
beyond $z=1$, the estimated 
median redshift of the FIRST survey (Cress \& Kamionkowski 1997).
If complete redshift information were acquired for such a sample,
the resulting data set
would, more 
reliably than data with mixed optical and radio information,
further our understanding of the intrinsic properties and evolution
of double-lobed radio sources, the behavior of the IGM density as a function of redshift, 
and the quasar-radio galaxy unification issue, and, perhaps most importantly,
be instrumental in determining the potential impact of angular size-redshift
studies in cosmology.
 
\acknowledgements

We wish to thank Chelsea T. Wald for her help in compiling the data sample,
as well as Alexandre Refregier, David Schminovich, Catherine Cress,
Jacqueline Van Gorkom, and Kevin H. Prendergast for their numerous 
insightful comments and suggestions. 
We acknowledge support from the NSF (grants AST-94-19906 and AST-94-21178), 
the IGPP/LLNL (DOE contract W-7405-ENG-48),
the STScI, the National Geographic Society (grant NGS No. 5393-094), Columbia University,
and Sun Microsystems.
This research has made use of the NASA/IPAC
Extragalactic Database (NED) which is operated by the Jet Propulsion Laboratory,
California Institute of Technology, under contract with the National Aeronautics
and Space Administration.

\clearpage

\footnotesize
\begin{planotable}{ccrrrrrc}
%\tablewidth{33pc}
\tablecaption{Double-Lobed Radio Quasar Data}
\tablehead{
\colhead{$\alpha$ (J2000)}         &
\colhead{$\delta$ (J2000)}         & \colhead{$z$}  &
\colhead{$\theta$ ($^{\arcsec}$)}  & \colhead{$S_t$ (mJy)}    &
\colhead{$S_l$ (mJy)}  & \colhead{$R$} & \colhead{Ref. Code}}
\startdata
00 22 44.3 & -01 45 51 & 0.691 &  85.7 &   242.2 &   232.7 &  0.04 & 1 \nl
00 41 25.9 & -01 43 15 & 1.679 &  19.3 &  1042.1 &  1042.1 &  0.00 & 1 \nl
01 03 29.4 & +00 40 54 & 1.436 &  26.3 &   114.7 &   114.7 &  0.00 & 1 \nl
01 19 10.0 & +01 31 28 & 0.520 &  84.2 &    60.3 &    58.1 &  0.04 & 4 \nl
01 33 52.7 & +01 13 46 & 1.370 & 100.5 &   109.9 &    93.2 &  0.18 & 1 \nl
02 10 08.5 & +01 18 39 & 0.870 & 154.4 &    33.6 &    27.0 &  0.24 & 4 \nl
02 25 07.9 & -00 35 32 & 0.687 &  12.5 &  1141.4 &  1141.4 &  0.00 & 1 \nl
02 39 13.6 & -01 18 16 & 1.794 &  15.1 &   237.5 &   237.5 &  0.00 & 1 \nl
02 45 34.0 & +01 08 13 & 1.520 &  52.5 &   330.1 &   318.4 &  0.04 & 1 \nl
02 50 48.6 & +00 02 08 & 0.766 &  16.5 &   111.1 &    91.3 &  0.22 & 1 \nl
03 15 42.4 & -01 51 23 & 1.480 &  27.9 &   278.2\tablenotemark{F} &   156.7 &  0.78 & 4 \nl
07 43 45.0 & +23 28 39 & 0.770 &  23.6 &   335.1 &   181.1 &  0.85 & 1 \nl
07 45 41.6 & +31 42 56 & 0.461 & 115.1 &  1454.7 &   840.0 &  0.73 & 1 \nl
07 52 28.7 & +37 50 52 & 1.200 &  28.2 &   395.7 &   365.9 &  0.08 & 1 \nl
07 53 28.3 & +33 50 52 & 2.070 &  27.3 &   150.3 &    87.2 &  0.72 & 1 \nl
08 02 20.5 & +30 35 43 & 1.640 &  53.2 &    67.5 &    39.7 &  0.70 & 3 \nl
08 09 06.2 & +29 12 35 & 1.470 & 131.4 &   312.9 &   291.4 &  0.07 & 3 \nl
08 11 36.9 & +28 45 03 & 1.910 &  58.6 &   102.2 &    62.8 &  0.63 & 1 \nl
08 14 09.3 & +32 37 31 & 0.842 &  24.2 &   514.5 &   387.6 &  0.33 & 1 \nl
08 14 30.6 & +38 58 35 & 2.621 &  24.7 &    72.3 &    72.3 &  0.00 & 1 \nl
08 17 35.1 & +22 37 17 & 0.980 &  23.6 &  1315.3 &  1150.5 &  0.14 & 1 \nl
08 17 40.2 & +34 54 52 & 1.348 &  52.8 &    26.2 &    19.9 &  0.32 & 2 \nl
08 28 06.8 & +39 35 40 & 0.762 &  64.6 &    86.7 &    81.6 &  0.06 & 1 \nl
08 32 36.7 & +33 32 05 & 1.100 &  30.1 &   354.2 &   354.2 &  0.00 & 1 \nl
08 32 48.4 & +42 24 59 & 1.051 &  16.1 &   456.1 &   168.7 &  1.70 & 1 \nl
08 46 59.3 & +34 48 25 & 1.575 &  30.5 &   101.6 &    62.1 &  0.64 & 2 \nl
08 47 56.4 & +31 47 58 & 1.834 & 161.1 &  1589.7 &  1568.0 &  0.01 & 1 \nl
08 52 34.2 & +42 15 28 & 0.978 &  20.0 &   459.6 &   459.6 &  0.00 & 1 \nl
09 04 29.6 & +28 19 33 & 1.121 &  22.5 &   130.1 &    90.9 &  0.43 & 1 \nl
09 07 45.5 & +38 27 39 & 1.740 &  15.2 &   156.8 &   112.6 &  0.39 & 1 \nl
09 13 45.5 & +40 56 27 & 0.442 &  20.6 &    16.4 &     8.1 &  1.02 & 1 \nl
09 13 52.4 & +39 02 12 & 0.638 &  53.0 &   145.2 &   145.2 &  0.00 & 1 \nl
09 21 46.6 & +37 54 10 & 1.108 &  51.9 &   825.0 &   547.0 &  0.51 & 1 \nl
09 25 54.7 & +40 04 14 & 0.470 & 262.4 &    77.4 &    68.2 &  0.13 & 4 \nl
09 31 52.8 & +34 39 20 & 2.304 &  12.8 &    26.6 &    18.8 &  0.42 & 2 \nl
09 37 04.0 & +29 37 04 & 0.450 & 157.4 &    29.4 &    26.8 &  0.10 & 3 \nl
09 41 04.1 & +38 53 51 & 0.618 &  51.1 &   668.6 &   443.8 &  0.51 & 1 \nl
09 52 31.9 & +35 12 53 & 1.875 &  25.7 &   339.0 &    34.3 &  8.88 & 1 \nl
09 55 48.1 & +35 33 23 & 1.241 &  18.8 &   522.7 &   522.7 &  0.00 & 1 \nl
09 58 02.8 & +38 29 58 & 1.394 &  18.9 &   432.8 &   432.8 &  0.00 & 1 \nl
10 00 21.8 & +22 33 19 & 0.419 &  34.1 &  1117.2 &  1117.2 &  0.00 & 1 \nl
10 04 45.8 & +22 25 19 & 0.974 &  66.4 &   578.9\tablenotemark{F} &   545.0 &  0.06 & 1 \nl
10 10 27.5 & +41 32 38 & 0.612 &  31.8 &  1734.8 &  1394.5 &  0.24 & 1 \nl
10 14 35.8 & +27 49 03 & 0.899 &  13.1 &   514.8 &   514.8 &  0.00 & 1 \nl
10 17 49.3 & +27 32 05 & 0.469 &  21.1 &  1318.7 &  1318.7 &  0.00 & 1 \nl
10 18 25.5 & +38 05 33 & 0.380 &  48.4 &   275.2 &   242.5 &  0.14 & 1 \nl
10 20 41.1 & +39 58 11 & 0.830 & 159.9 &     9.6\tablenotemark{F} &     9.6 &  0.00 & 4 \nl
10 21 17.5 & +34 37 23 & 1.400 &  18.3 &   457.5 &   144.2 &  2.17 & 1 \nl
10 51 29.4 & +23 48 02 & 1.274 &  15.4 &   485.2 &   485.2 &  0.00 & 1 \nl
10 52 50.1 & +33 55 05 & 1.405 &  32.7 &    21.5 &     8.7 &  1.47 & 4 \nl
11 03 13.3 & +30 14 43 & 0.380 &  73.1 &   167.5 &    59.8 &  1.80 & 1 \nl
11 07 26.8 & +36 16 12 & 0.393 &  20.4 &   611.9\tablenotemark{F} &   611.9 &  0.00 & 1 \nl
11 08 37.7 & +38 58 41 & 0.781 &  67.1 &   877.7 &   867.8 &  0.01 & 1 \nl
11 10 40.2 & +30 19 09 & 1.520 &  42.3 &    91.0 &    68.2 &  0.33 & 3 \nl
11 14 38.6 & +40 37 20 & 0.734 &  13.3 &  3037.2 &  3037.2 &  0.00 & 1 \nl
11 19 03.2 & +38 58 52 & 0.733 &  90.0 &   141.1 &   129.3 &  0.09 & 1 \nl
11 34 54.5 & +30 05 26 & 0.614 &  15.1 &  1147.4 &  1147.4 &  0.00 & 1 \nl
11 48 18.8 & +31 54 11 & 0.549 &  20.7 &    94.0 &    45.8 &  1.05 & 1 \nl
12 06 17.3 & +38 12 35 & 0.838 &  35.1 &   241.0 &   241.0 &  0.00 & 1 \nl
12 10 37.7 & +31 57 07 & 0.388 &  80.2 &   276.3 &   254.3 &  0.09 & 1 \nl
12 23 11.2 & +37 07 02 & 0.489 &  36.0 &   477.3 &   430.1 &  0.11 & 1 \nl
12 30 52.5 & +39 30 00 & 2.217 &  51.9 &   223.9 &   219.6 &  0.02 & 1 \nl
12 33 28.3 & +34 39 42 & 0.847 &  31.0 &    41.8 &    33.9 &  0.23 & 2 \nl
12 36 31.3 & +26 35 09 & 2.100 &  21.6 &   557.3 &   557.3 &  0.00 & 1 \nl
12 36 51.4 & +25 07 48 & 0.546 &  83.8 &   270.6 &   254.4 &  0.06 & 1 \nl
12 37 04.0 & +33 14 23 & 1.280 &  18.2 &   218.5 &    52.9 &  3.13 & 1 \nl
12 40 21.2 & +35 02 59 & 1.194 &  16.8 &   222.1 &   222.1 &  0.00 & 1 \nl
12 47 20.7 & +32 09 01 & 0.949 &  22.9 &   470.1 &   338.0 &  0.39 & 1 \nl
12 50 25.5 & +30 16 40 & 1.061 &  28.6 &   430.6 &   430.6 &  0.00 & 1 \nl
12 54 10.5 & +39 33 23 & 2.104 &  32.4 &    56.9 &    56.9 &  0.00 & 1 \nl
12 59 02.1 & +39 00 13 & 0.978 &  20.9 &   297.1\tablenotemark{F} &   258.1 &  0.15 & 1 \nl
13 00 33.4 & +40 09 07 & 1.659 &  19.8 &  1287.2\tablenotemark{F} &  1287.2 &  0.00 & 1 \nl
13 08 56.8 & +27 08 12 & 1.537 &  15.2 &   334.6 &   180.2 &  0.86 & 1 \nl
13 41 08.2 & +39 14 49 & 0.580 &  12.6 &    84.2 &    84.2 &  0.00 & 1 \nl
13 42 10.9 & +28 28 47 & 0.330 &  94.6 &   224.2 &   222.7 &  0.01 & 1 \nl
13 42 54.5 & +28 28 05 & 1.037 &  33.1 &   184.1 &   118.5 &  0.55 & 1 \nl
13 44 25.5 & +38 41 29 & 1.533 &  18.8 &   286.9 &   286.9 &  0.00 & 1 \nl
13 50 15.0 & +38 12 05 & 1.390 &  16.8 &   180.3 &   180.3 &  0.00 & 1 \nl
13 53 36.0 & +26 31 48 & 0.310 & 173.2 &   244.5 &   222.2 &  0.10 & 1 \nl
14 11 55.3 & +34 15 11 & 1.820 &  20.2 &   201.1 &    98.4 &  1.04 & 1 \nl
14 16 58.3 & +34 28 53 & 0.750 &  14.8 &   151.1 &   151.1 &  0.00 & 1 \nl
14 25 50.8 & +24 04 03 & 0.649 &  20.4 &  1479.5\tablenotemark{F} &  1158.6 &  0.28 & 1 \nl
14 27 35.7 & +26 32 14 & 0.366 & 228.4 &   368.5 &   314.4 &  0.17 & 1 \nl
14 37 56.5 & +35 19 37 & 0.540 &  14.3 &    89.4 &    89.4 &  0.00 & 1 \nl
14 46 26.8 & +41 33 18 & 0.675 & 102.9 &   534.1 &   531.1 &  0.01 & 1 \nl
15 14 43.0 & +36 50 50 & 0.370 &  54.8 &  1001.2 &   930.2 &  0.08 & 1 \nl
15 57 30.0 & +33 04 46 & 0.942 &  33.6 &   168.2 &    82.2 &  1.05 & 1 \nl
16 08 11.2 & +28 49 02 & 1.989 &  30.3 &   589.5 &   570.5 &  0.03 & 1 \nl
16 13 51.4 & +37 42 59 & 1.630 &  16.3 &   283.6 &   283.6 &  0.00 & 1 \nl
16 22 29.9 & +35 31 26 & 1.473 &  21.9 &   381.6\tablenotemark{F} &   351.2 &  0.09 & 1 \nl
16 24 22.1 & +39 24 42 & 1.120 &  21.3 &   254.2 &   121.9 &  1.09 & 1 \nl
16 24 39.3 & +23 45 12 & 0.927 &  22.7 &  2587.9\tablenotemark{F} &  2171.9 &  0.19 & 1 \nl
16 25 30.8 & +27 05 47 & 0.525 &  23.1 &   532.0 &   299.1 &  0.00 & 1 \nl
16 30 46.2 & +36 13 07 & 1.256 &  15.2 &   543.3 &   434.4 &  0.25 & 1 \nl
16 33 02.2 & +39 24 27 & 1.023 &  17.3 &    69.6 &    69.6 &  0.00 & 1 \nl
16 36 36.4 & +26 48 09 & 0.561 &  40.1 &  1337.9\tablenotemark{F} &  1337.9 &  0.00 & 1 \nl
17 03 07.7 & +37 51 25 & 2.450 &  19.1 &   111.1 &   111.1 &  0.00 & 1 \nl
17 06 48.1 & +32 14 22 & 1.070 &  53.0 &   136.2 &    99.9 &  0.36 & 4 \nl
21 35 13.1 & -00 52 43 & 2.660 &  58.4 &   324.4\tablenotemark{F} &   323.1 &  0.00 & 4 \nl
22 14 10.0 & +00 52 28 & 0.910 &  39.6 &   121.9\tablenotemark{F} &    91.8 &  0.33 & 1 \nl
23 36 24.1 & +00 02 46 & 1.100 &  59.8 &   227.1\tablenotemark{F} &   209.4 &  0.08 & 4 \nl
23 44 40.0 & -00 32 31 & 0.500 & 168.0 &    36.4\tablenotemark{F} &    18.0 &  1.02 & 4 \nl
23 47 24.5 & +00 52 44 & 0.400 &  19.1 &    94.8\tablenotemark{F} &    94.8 &  0.00 & 2   
\enddata
\tablecomments{Data for our sample of 103 FR-II quasars found within the currently available
region of the FIRST survey. The coordinates and redshifts of the quasars are taken from
the Veron-Cetty \& Veron (1996), Hewitt \& Burbidge (1993), and FIRST Bright QSO Survey
(Gregg \etal 1996; Becker \etal 1997) catalogs, with these four references respectively
denoted by reference codes 1, 2, 3, or 4 in column 8.
The peak-to-peak angular sizes, $\theta$, are measured directly from the FIRST data. The total
1.4 GHz flux densities, $S_t$, are taken from the NVSS survey, with a superscripted
F indicating that only FIRST fluxes were available for that source. The lobe flux densities,
$S_l$, are obtained by subtracting the FIRST flux of the core component (if any) from
$S_t$. The core-to-lobe flux density ratio is given by $R$.}
\label{tab1}
\end{planotable}
\normalsize

\clearpage

\tiny
\begin{planotable}{ccccccc|ccc}
%\tablewidth{45pc}
\tablecaption{Selected Results Of Parametric Fits For $c$, $\beta$, $x$, and $n$}
\tablehead{
\colhead{Model} & \colhead{$\Omega_0$}      & \colhead{Limit}     &
\colhead{$c$}      & \colhead{$\beta$}  &
\colhead{$x$}      & \colhead{$n$}    &
\colhead{$\beta_l$}      & \colhead{$x_l$}  &
\colhead{$n_l$}}
\startdata
1 & 1.00 & $l_{max}$ & $-0.828 \pm 0.379$ & $-0.168 \pm 0.067$ & $ 3.672 \pm 0.376$ & $-0.211 \pm 0.456$ & $-0.150 \pm 0.066$ & $ 3.793 \pm 0.404$ & $-0.260 \pm 0.457$ \nl
2 & 0.10 & $l_{max}$ & $-0.827 \pm 0.423$ & $-0.102 \pm 0.061$ & $ 4.513 \pm 0.352$ & $-0.366 \pm 0.507$ & $-0.182 \pm 0.059$ & $ 4.644 \pm 0.380$ & $ 0.017 \pm 0.509$ \nl
2 & 0.30 & $l_{max}$ & $-0.694 \pm 0.362$ & $-0.117 \pm 0.061$ & $ 3.993 \pm 0.350$ & $-0.227 \pm 0.439$ & $-0.197 \pm 0.067$ & $ 4.094 \pm 0.376$ & $ 0.114 \pm 0.459$ \nl
2 & 0.90 & $l_{max}$ & $-0.808 \pm 0.378$ & $-0.166 \pm 0.067$ & $ 3.712 \pm 0.375$ & $-0.193 \pm 0.457$ & $-0.148 \pm 0.065$ & $ 3.835 \pm 0.403$ & $-0.242 \pm 0.458$ \nl
3 & 0.10 & $l_{max}$ & $-0.853 \pm 0.427$ & $-0.121 \pm 0.061$ & $ 4.280 \pm 0.364$ & $-0.335 \pm 0.502$ & $-0.182 \pm 0.062$ & $ 4.399 \pm 0.392$ & $-0.054 \pm 0.512$ \nl
3 & 0.30 & $l_{max}$ & $-0.678 \pm 0.366$ & $-0.113 \pm 0.063$ & $ 4.087 \pm 0.357$ & $-0.217 \pm 0.449$ & $-0.167 \pm 0.062$ & $ 4.198 \pm 0.382$ & $ 0.024 \pm 0.454$ \nl
3 & 0.90 & $l_{max}$ & $-0.800 \pm 0.379$ & $-0.164 \pm 0.067$ & $ 3.725 \pm 0.376$ & $-0.187 \pm 0.457$ & $-0.147 \pm 0.065$ & $ 3.847 \pm 0.404$ & $-0.236 \pm 0.458$ \nl \hline
1 & 1.00 & $l_{*}$ & $-0.142 \pm 0.225$ & $-0.098 \pm 0.044$ & $ 2.843 \pm 0.445$ & $ 0.137 \pm 0.262$ & $-0.090 \pm 0.041$ & $ 2.884 \pm 0.496$ & $ 0.117 \pm 0.258$ \nl
2 & 0.10 & $l_{*}$ & $-0.141 \pm 0.221$ & $-0.064 \pm 0.042$ & $ 3.462 \pm 0.442$ & $ 0.082 \pm 0.266$ & $-0.067 \pm 0.039$ & $ 3.424 \pm 0.484$ & $ 0.090 \pm 0.260$ \nl
2 & 0.30 & $l_{*}$ & $-0.108 \pm 0.232$ & $-0.062 \pm 0.044$ & $ 3.204 \pm 0.433$ & $ 0.092 \pm 0.272$ & $-0.063 \pm 0.041$ & $ 3.231 \pm 0.477$ & $ 0.094 \pm 0.268$ \nl
2 & 0.90 & $l_{*}$ & $-0.122 \pm 0.225$ & $-0.095 \pm 0.044$ & $ 2.878 \pm 0.444$ & $ 0.152 \pm 0.261$ & $-0.088 \pm 0.041$ & $ 2.921 \pm 0.495$ & $ 0.134 \pm 0.258$ \nl
3 & 0.10 & $l_{*}$ & $-0.168 \pm 0.232$ & $-0.068 \pm 0.043$ & $ 3.332 \pm 0.449$ & $ 0.058 \pm 0.274$ & $-0.071 \pm 0.039$ & $ 3.351 \pm 0.492$ & $ 0.070 \pm 0.269$ \nl
3 & 0.30 & $l_{*}$ & $-0.153 \pm 0.230$ & $-0.065 \pm 0.042$ & $ 3.679 \pm 0.389$ & $ 0.084 \pm 0.277$ & $-0.066 \pm 0.040$ & $ 3.650 \pm 0.450$ & $ 0.088 \pm 0.273$ \nl
3 & 0.90 & $l_{*}$ & $-0.114 \pm 0.225$ & $-0.094 \pm 0.044$ & $ 2.895 \pm 0.445$ & $ 0.159 \pm 0.262$ & $-0.087 \pm 0.041$ & $ 2.938 \pm 0.496$ & $ 0.141 \pm 0.259$ 
\enddata
\tablecomments{The values for $\beta$, $x$, and $n$ are obtained using the total power, $P$,
while those for $\beta_l$, $x_l$, and $n_l$ are obtained using only the lobe power, $P_l$. Note
that these two sets of values agree to within the 1$\sigma$ errors listed. A limit
of $l_{max}$ means that all data points above the $\theta$-$z$ curve corresponding to $l_{min}$
for the given model were used, while a limit of $l_*$ means that 
points lying above the $\theta$-$z$ curve with a minimum at $65^{\arcsec}$ 
for that choice of $\Omega_0$ were rejected, corresponding to the subsample with $c \approx 0$.}
\label{tab2}
\end{planotable}
\normalsize

\clearpage

\tiny
\begin{planotable}{cccccccc}
%\tablewidth{8in}
\tablecaption{Selected Results Of Non-Parametric Analysis Between $s$, $P$, and $z$}
\tablehead{
\colhead{Model} & \colhead{$\Omega_0$}      & \colhead{Limit}     &
\colhead{$r_{sz}$}      & \colhead{$r_{sP}$}  &
\colhead{$r_{sP_{l}}$}    &
\colhead{$r_{Pz}$}      & \colhead{$r_{sz,P}$}}
\startdata
1 & 1.00 & $l_{max}$ & -0.196(0.051) & -0.343(0.001) & -0.318(0.002) &  0.693($<10^{-11}$) &  0.062(0.536) \nl
2 & 0.10 & $l_{max}$ & -0.199(0.057) & -0.329(0.002) & -0.302(0.004) &  0.749($<10^{-12}$) &  0.076(0.462) \nl
2 & 0.30 & $l_{max}$ & -0.157(0.120) & -0.277(0.006) & -0.258(0.011) &  0.726($<10^{-12}$) &  0.067(0.507) \nl
2 & 0.90 & $l_{max}$ & -0.188(0.062) & -0.337(0.001) & -0.312(0.002) &  0.697($<10^{-11}$) &  0.069(0.488) \nl
3 & 0.10 & $l_{max}$ & -0.205(0.048) & -0.314(0.002) & -0.284(0.006) &  0.746($<10^{-12}$) &  0.048(0.645) \nl
3 & 0.30 & $l_{max}$ & -0.176(0.084) & -0.271(0.008) & -0.243(0.017) &  0.739($<10^{-12}$) &  0.036(0.721) \nl
3 & 0.90 & $l_{max}$ & -0.183(0.069) & -0.335(0.001) & -0.311(0.002) &  0.697($<10^{-11}$) &  0.074(0.458) \nl \hline
1 & 1.00 & $l_{*}$ & -0.017(0.874) & -0.211(0.056) & - &  0.634($<10^{-8}$) & - \nl
2 & 0.10 & $l_{*}$ & -0.059(0.600) & -0.203(0.072) & - &  0.701($<10^{-9}$) & - \nl
2 & 0.30 & $l_{*}$ & -0.017(0.876) & -0.144(0.186) & - &  0.683($<10^{-9}$) & - \nl
2 & 0.90 & $l_{*}$ & -0.006(0.957) & -0.200(0.070) & - &  0.637($<10^{-8}$) & - \nl
3 & 0.10 & $l_{*}$ & -0.073(0.514) & -0.194(0.082) & - &  0.704($<10^{-9}$) & - \nl
3 & 0.30 & $l_{*}$ & -0.038(0.729) & -0.133(0.228) & - &  0.699($<10^{-9}$) & - \nl
3 & 0.90 & $l_{*}$ &  0.001(0.993) & -0.197(0.074) & - &  0.638($<10^{-8}$) & - 
\enddata
\tablecomments{The quantities $r_{ab}$ and $r_{ab,c}$ respectively connote the Spearman
rank correlation and partial rank correlation coefficients between quantities $a$, $b$, and $c$.
In each case, the number in parenthesis denotes the two-sided probability that a 
random data set could achieve
the associated value of $\abso{r}$ and thus gives the significance of the result.}
\label{tab3}
\end{planotable}
\normalsize

\clearpage

\tiny
\begin{planotable}{cccccc}
%\tablewidth{8in}
\tablecaption{Selected Results Of Non-Parametric Correlations with $R$}
\tablehead{
\colhead{Model} & \colhead{$\Omega_0$}      & \colhead{Limit}     &
\colhead{$r_{Rs}$}  &
\colhead{$r_{RP}$}      & 
\colhead{$r_{RP_{c}}$}}
\startdata
1 & 1.00 & $l_{max}$ & -0.501($<10^{-4}$) & -0.052(0.671) &  0.580($<10^{-5}$) \nl
2 & 0.10 & $l_{max}$ & -0.474($<10^{-3}$) & -0.006(0.964) &  0.588($<10^{-5}$) \nl
2 & 0.30 & $l_{max}$ & -0.503($<10^{-4}$) & -0.046(0.705) &  0.567($<10^{-5}$) \nl
2 & 0.90 & $l_{max}$ & -0.506($<10^{-4}$) & -0.053(0.669) &  0.580($<10^{-5}$) \nl
3 & 0.10 & $l_{max}$ & -0.503($<10^{-4}$) & -0.038(0.759) &  0.547($<10^{-5}$) \nl
3 & 0.30 & $l_{max}$ & -0.505($<10^{-4}$) & -0.042(0.732) &  0.558($<10^{-5}$) \nl
3 & 0.90 & $l_{max}$ & -0.507($<10^{-4}$) & -0.053(0.669) &  0.580($<10^{-5}$) \nl \hline
1 & 1.00 & $l_{*}$ & -0.419(0.003) & -0.125(0.377) &  0.558($<10^{-4}$) \nl
2 & 0.10 & $l_{*}$ & -0.463(0.001) &  0.004(0.979) &  0.631($<10^{-5}$) \nl
2 & 0.30 & $l_{*}$ & -0.484($<10^{-3}$) & -0.067(0.623) &  0.581($<10^{-4}$) \nl
2 & 0.90 & $l_{*}$ & -0.427(0.003) & -0.123(0.386) &  0.562($<10^{-4}$) \nl
3 & 0.10 & $l_{*}$ & -0.506($<10^{-3}$) & -0.038(0.776) &  0.573($<10^{-4}$) \nl
3 & 0.30 & $l_{*}$ & -0.486($<10^{-3}$) & -0.062(0.647) &  0.571($<10^{-4}$) \nl
3 & 0.90 & $l_{*}$ & -0.428(0.002) & -0.123(0.386) &  0.562($<10^{-4}$) 
\enddata
%\tablecomments{}
\label{tab4}
\end{planotable}
\normalsize

\clearpage

\begin{planotable}{cccccc}
%\tablewidth{8in}
\tablecaption{Best Fit Parameters To The $\theta$-$z$ Data}
\tablehead{
\colhead{Model} & \colhead{Limit}     &
\colhead{$\chi^2$ ($N,\nu,1-p$)}  &
\colhead{$a$}      & \colhead{$c$}      & 
\colhead{$\Omega_0$}}
\startdata
1 & $l_{max}$ & 1.776 (100,3,0.620) & 0.32 (0.19,0.55) & -0.83 (-1.64,-0.12) & 1.00 \nl
2 & $l_{max}$ & 0.822 (98,2,0.663) & 0.35 (0.21,0.57) & -0.64 (-1.36,-0.12) & 0.38 (0.28,1.00) \nl
3 & $l_{max}$ & 1.021 (97,2,0.600) & 0.32 (0.21,0.50) & -0.61 (-1.37,-0.05) & 0.35 (0.25,1.00) \nl 
Euclidean & -&  6.959 (103,4,0.138) & 0.57 (0.52,0.62) & - & - \nl \hline
1 & $l_{*}$ & 0.408 (83,3,0.939) & 0.15 (0.10,0.21) & -0.25 (-0.73,-0.29) & 1.00 \nl
2 & $l_{*}$ & 0.362 (83,2,0.834) & 0.15 (0.10,0.35) & -0.19 (-0.67,-0.31) & 0.84 (0.03,1.00) \nl
3 & $l_{*}$ & 0.408 (83,2,0.816) & 0.15 (0.10,0.29) & -0.23 (-0.66,-0.29) & 0.93 (0.00,1.00) \nl
\enddata
\tablecomments{$N$ denotes the number of points falling within the angular size
cutoffs in each case, which is then divided into five bins.
For normally-distributed data (which ours are not)
the value of $1-p$ would give the significance of the result,
where $p$ is the cumulative distribution function for the $\chi^2$
probability function with $\nu$ degrees of freedom. Each best-fit parameter value is accompanied
by the 1$\sigma$ confidence limits in parentheses.}
\label{tab5}
\end{planotable}

\clearpage

\begin{planotable}{ccccc}
%\tablewidth{8in}
\tablecaption{Results For $H_0$ in km s$^{-1}$ Mpc$^{-1}$ With 1$\sigma$ Confidence Limits}
\tablehead{
\colhead{Model} & \colhead{Limit}     &
\colhead{$H_0$ [$\phi_u = 45^{\circ}$]}      & 
\colhead{$H_0$ [$\phi_u = 90^{\circ}$]}}
\startdata
1 & $l_{max}$ & 99 (59,170) & 62 (37,107) \nl
2 & $l_{max}$ & 107 (64,174) & 68 (41,110) \nl
3 & $l_{max}$ & 98 (64,152) & 62 (40,96) \nl \hline
1 & $l_{*}$ & 55 (36,77) & 39 (26,54) \nl
2 & $l_{*}$ & 49 (33,114) & 35 (23,81) \nl
3 & $l_{*}$ & 48 (32,93) & 34 (23,66) 
\enddata
%\tablecomments{}
\label{tab6}
\end{planotable}

\clearpage

\figcaption[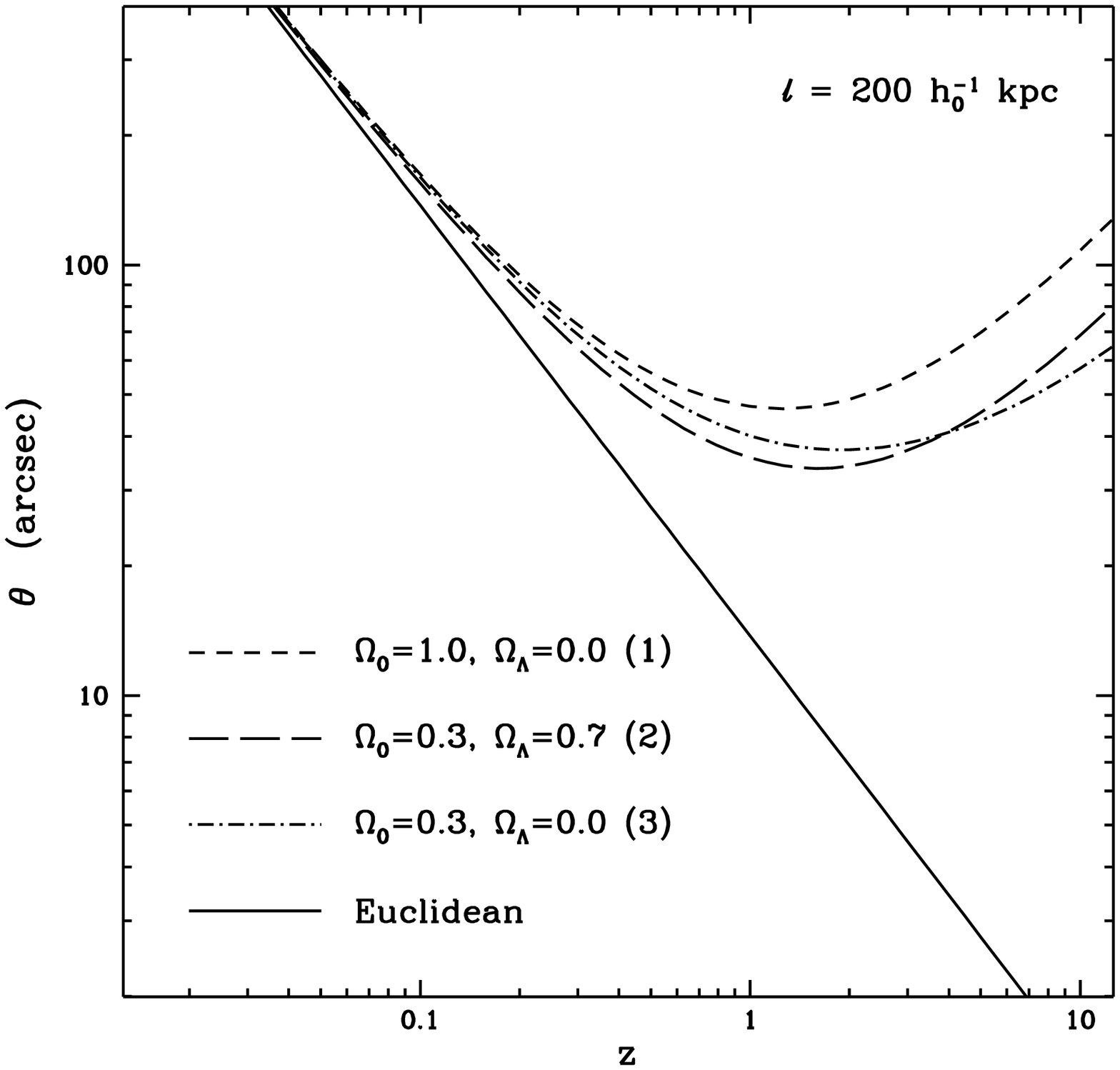]{The $\theta$-$z$ relation for de-projected rods
of length 200$h_{0}^{-1}$ kpc for different cosmologies. The choices for
$\Omega_0$ and $\Omega_{\Lambda}$ in the three Friedmann models (denoted
in parentheses) are listed on the Figure. The curve for a static, Euclidean universe is shown
for comparison. In practice, the curves actually define upper limits to the observed
angular sizes, since projection effects will scatter the observed
sizes downward. Note the presence of the minimum near $z \sim 1.5$ in the
Friedmann models.
\label{fig1}
}

\figcaption[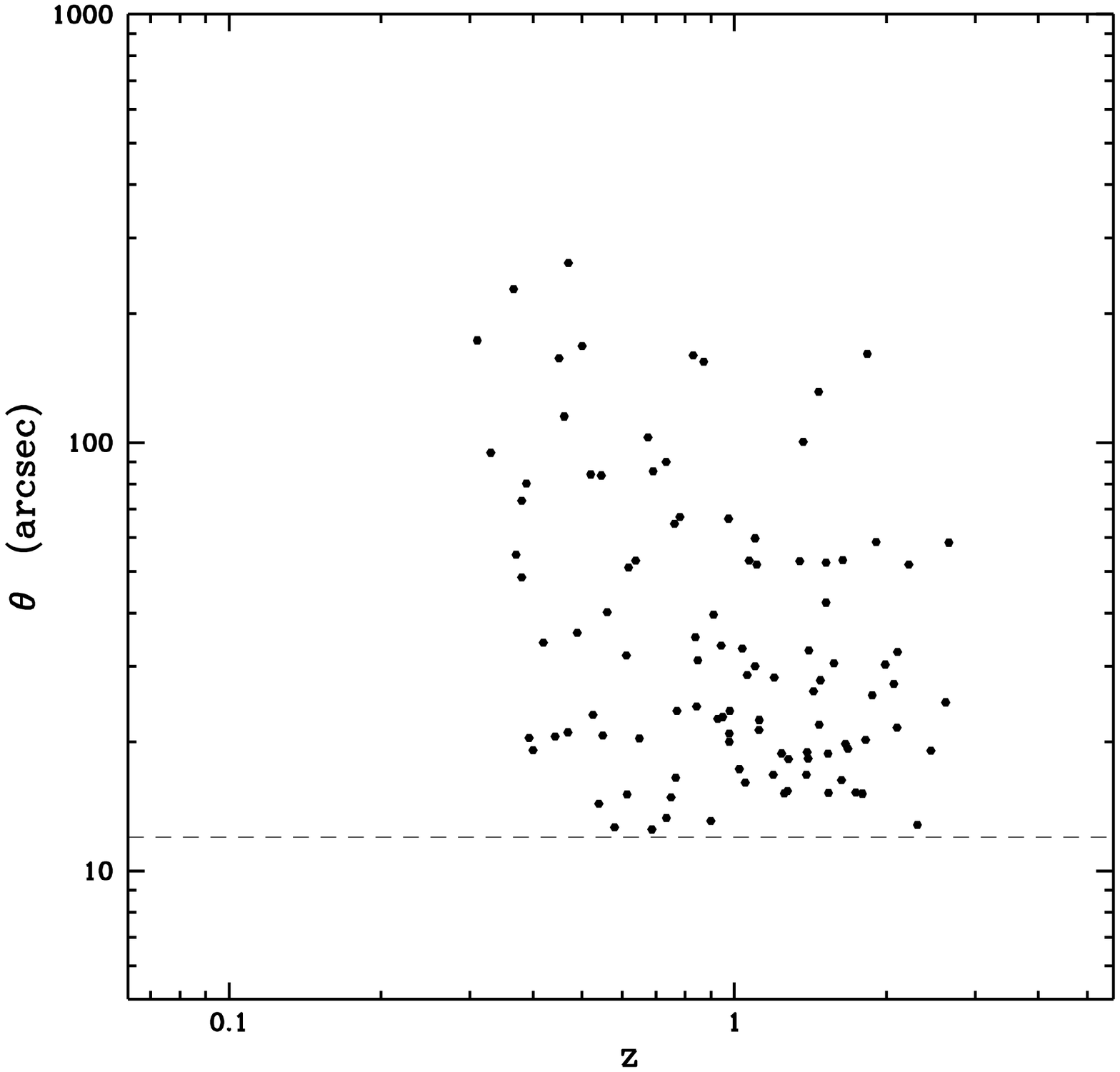]{Scatter plot of the peak-to-peak angular sizes, $\theta$, vs. redshift.
The dashed line represents the effective resolution limit at $12^{\arcsec}$, below which accurate 
morphological classifications could not be determined.
\label{fig2}
}

\figcaption[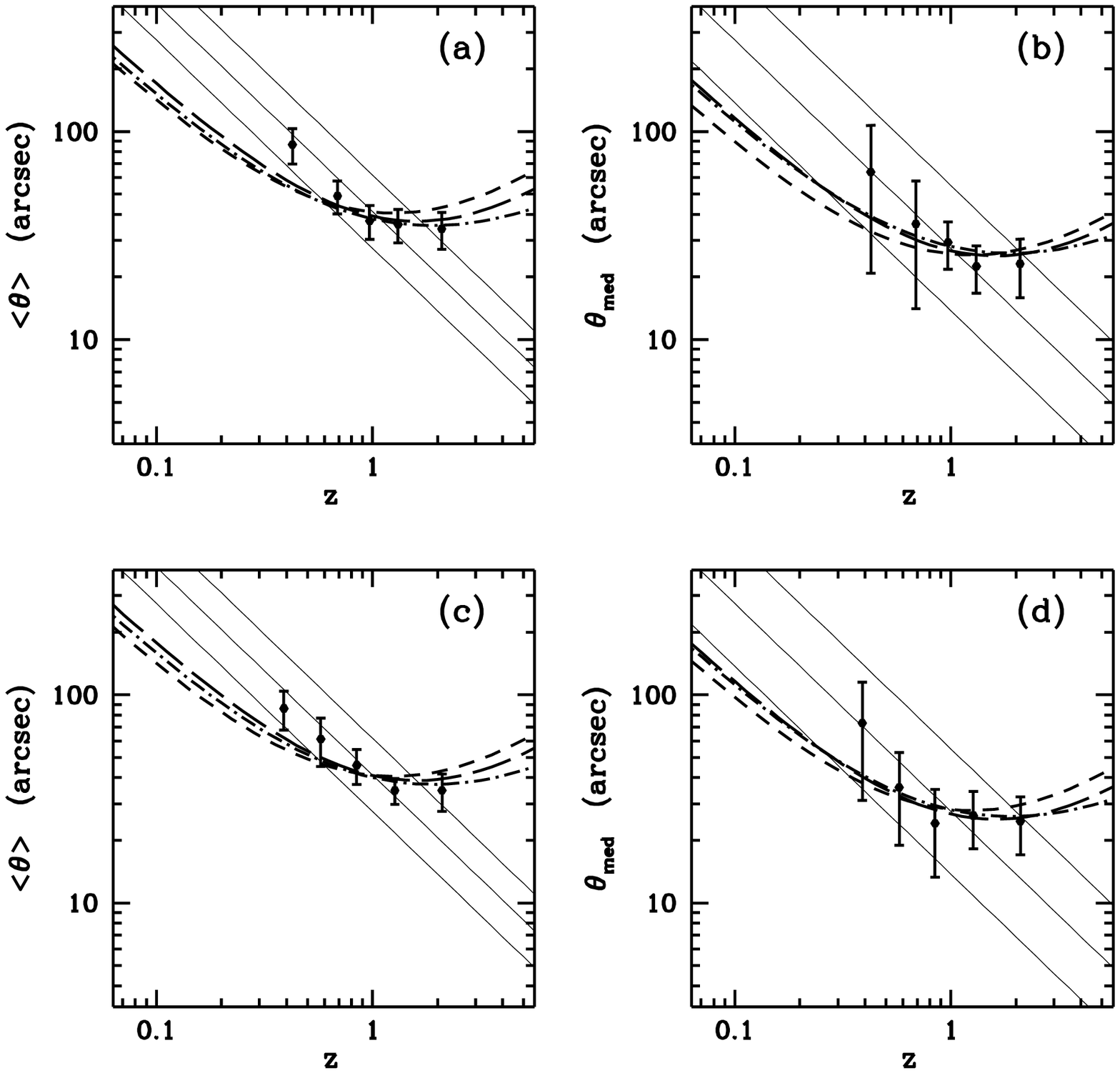]{Central values of $\theta$ vs. redshift using different analytical
techniques. In all plots, the short-dashed, long-dashed, and dot-dashed curves are the
respective predictions of Friedmann models 1, 2, and 3, assuming the density parameter
values listed in Figure 3a, and the thin, solid lines represent Euclidean curves
($\theta \propto z^{-1}$). The curves shown are not the best-fit results, but merely
visual estimates intended to provide a template for comparison. Note that in all cases,
the observed data are generally consistent with curvature and not with Euclidean
models. (a) Mean angular size $\VEV{\theta}$, binned in redshift with roughly equal
numbers per bin. The error bars represent the standard errors in the mean values. (b) Median
angular sizes, $\theta_{med}$, binned in redshift with roughly equal numbers per bin. The
error bars represent the median absolute deviation in each bin. (c) $\VEV{\theta}$ binned
in equal intervals of $(1+z)^{-3/2}$, which corresponds to equal time per bin in an
Einstein-de Sitter universe. The error bars represent the standard errors in the mean values.
(d) $\theta_{med}$ binned in equal intervals of $(1+z)^{-3/2}$. The
error bars represent the median absolute deviation in each bin.
\label{fig3}
}

\figcaption[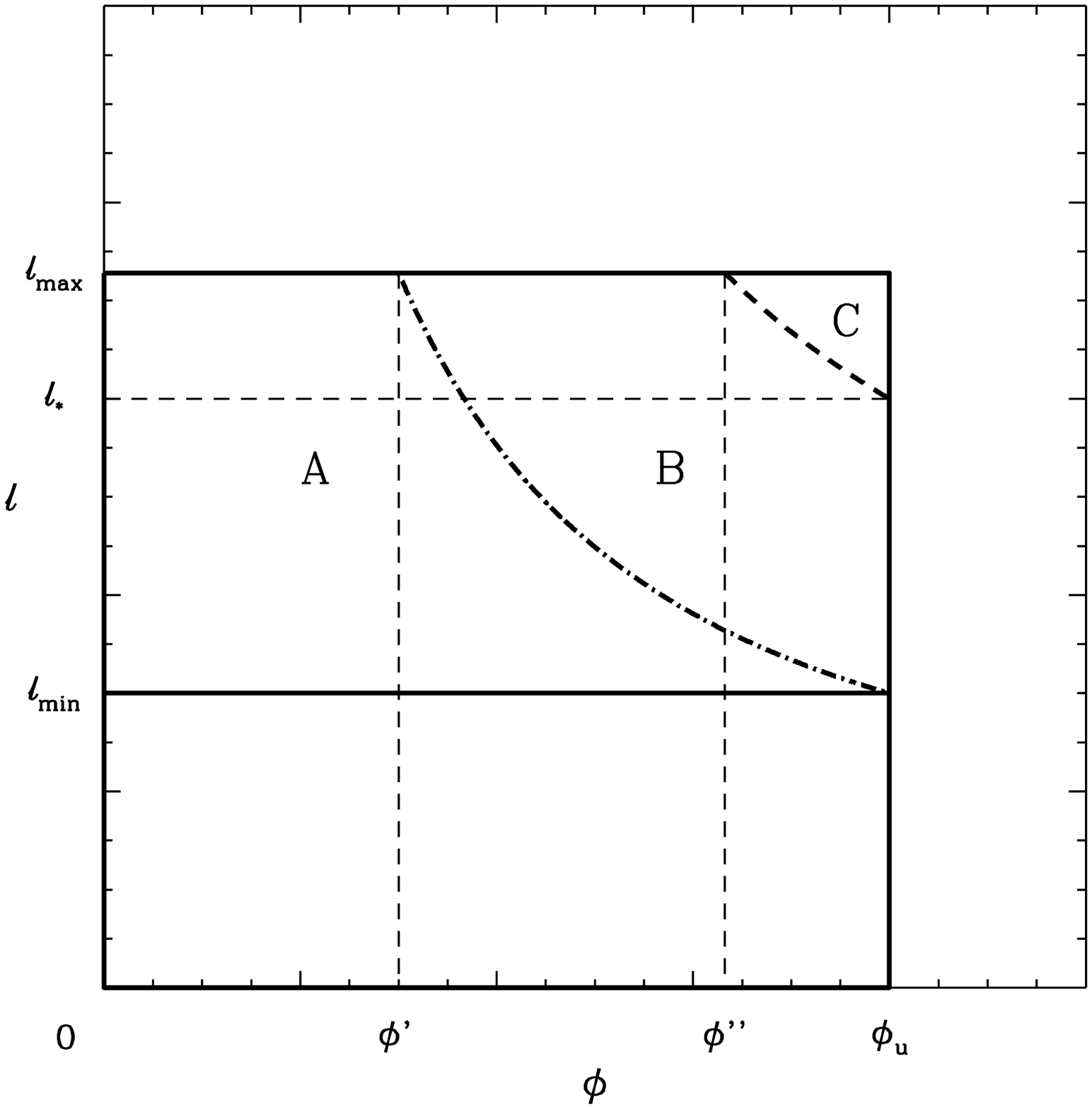]{The parameter space defined by the intrinsic sizes, $l$ (ordinate),
and projection angles with respect to the line of sight, $\phi$ (abscissa), of FR-II
quasars. The values of $l$ can range from 0 to an arbitrary $l_{max}$, while $\phi$ ranges
from 0 to some upper limit, $\phi_u$, which would correspond to $90^{\circ}$ for randomly
oriented quasars, or $\sim 45^{\circ}$ according to unification models, so that the 
largest bold rectangle 
represents the accessible portion of the space. For a survey with a given resolution limit,
we define an intrinsic size, $l_{min}$, so that maximally de-projected sources with this
size correspond to the smallest angular scale at which morphologies can be accurately
determined. Any objects intrinsically larger than $l_{min}$ can then be accurately classified
if their projection angles exceed some critical value, given by the long-dashed line (e.g.,
objects with $l=l_{max}$ can be accurately classified with $\phi$ ranging down to $\phi^{\prime}$).
Thus, the combined areas of regions $B$ and $C$ define the subspace of an 
accurately classified sample of sources with $l_{min} < l < l_{max}$ from a single radio survey.
In general, one can introduce an upper limit $l_u = l_* < l_{max}$ which will 
exclude objects larger than $l_*$ if their projection angles lie to the
right of the bold, dashed line, thus limiting the sample
to region $B$.
\label{fig4}
}

\figcaption[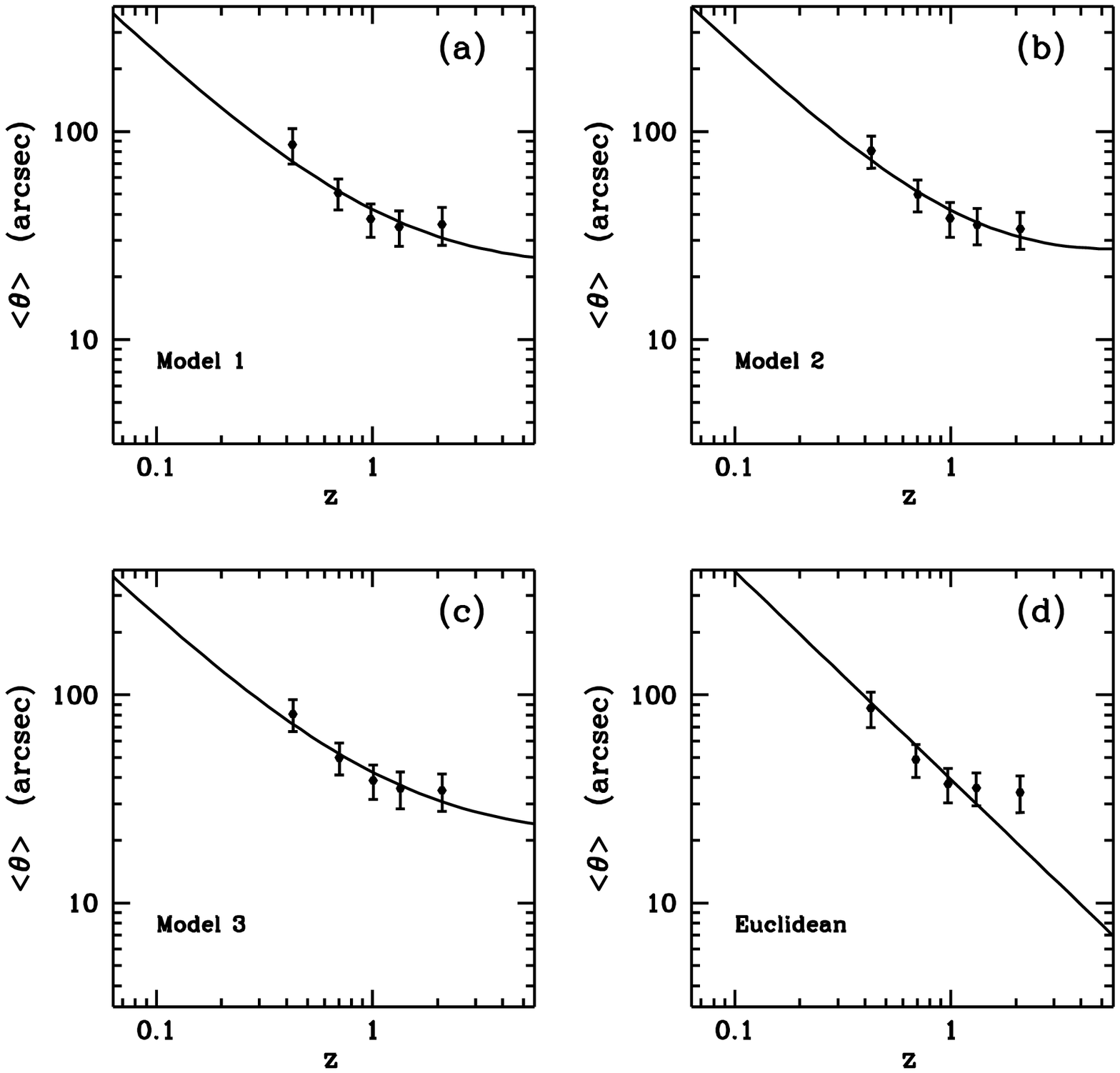]{Best-fit curves to the $\theta$-$z$ data with $l_u = l_{max}$
for the various models explored. (a) Model 1 with $a = 0.32$, $c = -0.83$, and $\Omega_0 = 1$.
(b) Model 2 with $a = 0.35$, $c = -0.64$, and $\Omega_0 = 0.38$.
(c) Model 3 with $a = 0.32$, $c = -0.61$, and $\Omega_0 = 0.35$.
(d) Euclidean model with $a = 0.57$.
Note that the Friedmann models appear to fit the data equally well, while the Euclidean
model constitutes a relatively poor fit. 
\label{fig5}
}

\figcaption[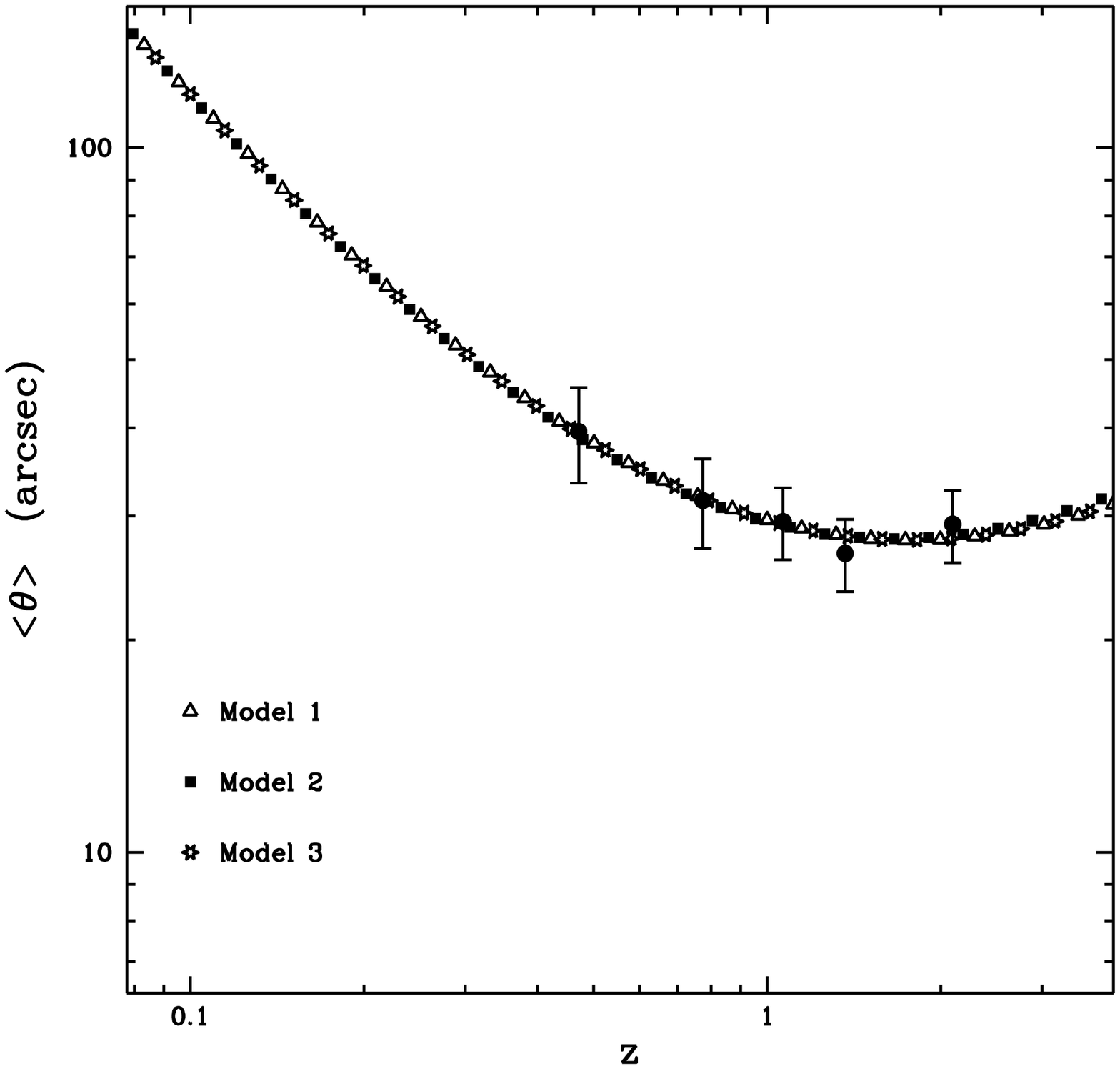]{Best-fit results obtained with $l_u = l_{*}$ for models 1, 2, and 3.
These Friedmann models all yield nearly identical values for $a$, $c$, and $\Omega_0$,
so that their $\theta$-$z$ curves, traced by the different point styles in the Figure, virtually
overlap.
\label{fig6}
}

\clearpage

\plotone{fig1.ps}

\clearpage

\plotone{fig2.ps}

\clearpage

\plotone{fig3.ps}

\clearpage

\plotone{fig4.ps}

\clearpage

\plotone{fig5.ps}

\clearpage

\plotone{fig6.ps}

\end{document}